%% file: main.tex
\documentclass[INR,biber]{nowfnt} 

\usepackage[utf8]{inputenc}
\usepackage{booktabs}
\usepackage[skip=0pt]{caption}
\usepackage[inline]{enumitem}
\usepackage{acronym}
\usepackage{multirow}
\usepackage{subfigure}
\usepackage{dsfont}
\usepackage{balance}
\usepackage{mathtools}
\usepackage{tcolorbox}
\usepackage{algorithm}
\usepackage{algpseudocode}
\usepackage{amsmath}
\usepackage{graphics}
\usepackage{epsfig}
\usepackage[figuresright]{rotating}
\usepackage{tablefootnote}
\usepackage{mwe}
\usepackage{lipsum}

\acrodef{IR}{Information Retrieval}
\acrodef{PTM}{Pre-training Method}
\acrodef{BERT}{Bidirectional Encoder Representations from Transformers}
\acrodef{NLP}{Nature Language Processing}

\title{Pre-training Methods in Information Retrieval}

\booltrue{authortwocolumn}
\maintitleauthorlist{ 
Yixing Fan \\
ICT, CAS, China \\
fanyixing@ict.ac.cn
\and
Xiaohui Xie \\
Tsinghua University \\
xiexiaohui@mail.tsinghua.edu.cn
\and
Yinqiong Cai \\
ICT, CAS, China \\
caiyinqiong18s@ict.ac.cn
\and
Jia Chen \\
Tsinghua University  \\
chenjia0831@gmail.com
\and
Xinyu Ma \\
ICT, CAS, China\\
maxinyu17g@ict.ac.cn
\and
Xiangsheng Li \\
Tsinghua University  \\
lixsh6@gmail.com
\and
Ruqing Zhang \\
ICT, CAS, China\\
zhangruqing@ict.ac.cn
\and
Jiafeng Guo \\
ICT, CAS, China\\
guojiafeng@ict.ac.cn
}

\issuesetup
{%
 volume        = xx,
 issue         = xx,
 pubyear       = 2021,
 isbn          = xxx-x-xxxxx-xxx-x,
 eisbn         = xxx-x-xxxxx-xxx-x,
 doi           = 10.1561/XXXXXXXXX,
 firstpage     = 1, 
 lastpage      = 18
 }

\author[1]{Yixing Fan$^*$}
\affil[1]{ICT, CAS, China; fanyixing@ict.ac.cn}
\author[2]{Xiaohui Xie$^*$}
\affil[2]{Tsinghua University; xiexiaohui@mail.tsinghua.edu.cn}
\author[1]{Yinqiong Cai}
\affil[1]{ICT, CAS, China; caiyinqiong18s@ict.ac.cn}
\author[2]{Jia Chen}
\affil[2]{Tsinghua University; chenjia0831@gmail.com}
\author[1]{Xinyu Ma}
\affil[1]{ICT, CAS, China; maxinyu17g@ict.ac.cn}
\author[2]{Xiangsheng Li}
\affil[2]{Tsinghua University; lixsh6@gmail.com}
\author[1]{Ruqing Zhang}
\affil[1]{ICT, CAS, China; zhangruqing@ict.ac.cn}
\author[1]{Jiafeng Guo$^\star$}
\affil[1]{ICT, CAS, China; guojiafeng@ict.ac.cn}

\articledatabox{\nowfntstandardcitation}

\begin{document}

\newcommand\blfootnote[1]{%
\begingroup
\renewcommand\thefootnote{}\footnote{#1}%
\addtocounter{footnote}{-1}%
\endgroup
}
\blfootnote{$^{*}$ Yixing Fan and Xiaohui Xie contributed equally.}
\blfootnote{$\star$ Corresponding authors.}

\makeabstracttitle

\begin{abstract}
The core of information retrieval (IR) is to identify relevant information from large-scale resources and return it as a ranked list to respond to the user's information need.
In recent years, the resurgence of deep learning has greatly advanced this field and leads to a hot topic named NeuIR (i.e., neural information retrieval), especially the paradigm of pre-training methods (PTMs). Owing to sophisticated pre-training objectives and huge model size, pre-trained models can learn universal language representations from massive textual data, which are beneficial to the ranking task of IR. Recently, a large number of works, which are dedicated to the application of PTMs in IR, have been introduced to promote the retrieval performance. 
Considering the rapid progress of this direction, this survey aims to provide a systematic review of pre-training methods in IR. To be specific, we present an overview of PTMs applied in different components of an IR system, including the retrieval component, the re-ranking component, and other components. In addition, we also introduce PTMs specifically designed for IR, and summarize available datasets as well as benchmark leaderboards. 
Moreover, we discuss some open challenges and highlight several promising directions, with the hope of inspiring and facilitating more works on these topics for future research.
\end{abstract}

\input{Sections/1-introduction}
\input{Sections/2-background}
\input{Sections/3-retrieval}
\input{Sections/4-reranking}
\input{Sections/5-othercomponent}
\input{Sections/6-pretrainingforIR}
\input{Sections/7-resource}
\input{Sections/8-challenge}

\chapter{Conclusion}
In this paper, we present a comprehensive overview of \acp{PTM} in IR, and gain some insights for future development. 
It includes the background of IR, a detailed description of PTMs applied in different components of IR, and a summary of related resources. 
Specifically, we describe the concepts of IR in a hierarchical view, and review the major paradigms of each stage. 
Then we thoroughly survey \acp{PTM} applied in different components of IR systems, including the first-stage retrieval component, the re-ranking component, and other components. 
In addition, we describe works in designing novel \acp{PTM} tailored for IR. 
Finally, we highlight several challenges on this topic and discuss potential research directions in this area. 
We hope this survey can help researchers who are interested in \acp{PTM} in IR, and will motivate new ideas to further explore this promising field.

\begin{acknowledgements}
\end{acknowledgements}

\backmatter 

\printbibliography

\end{document}

%% file: Sections/1-introduction.tex

\chapter{Introduction}
\label{section:introduction}

Information retrieval (IR) is a fundamental task in many real-world applications, such as Web search, question answering systems, digital libraries, and so on. 
The core of IR is to identify information resources relevant to user's information need (e.g., query or question) from a large collection. 
Since there might be more than one relevant resource, the returned result is often organized as a ranked list of documents (e.g., Web pages, answers, or responses) according to their relevance degree against the information need. 
Such ranking property of IR makes it different from other tasks, and researchers have devoted substantial efforts to develop a variety of ranking models in IR.

Over the past decades, many different ranking models have been introduced and studied, including vector space models \citep{Salton1975}, probabilistic ranking models \citep{robertson1976relevance}, and learning to rank (LTR) models \citep{li2014learning}. 
These methods have been successfully applied in many different IR applications, such as Web search engines like Google, news recommender systems like Toutiao, community question answering platforms like Quora, to name a few. 
More recently, a large variety of neural ranking models have been proposed, leading to a hot topic named NeuIR~\citep{Craswell2016ReportOT} (i.e., neural information retrieval). 
Different from previous non-neural ranking models that rely on elaborately-designed features and manually-designed functions, neural ranking models can automatically learn low-level dense representations from data as ranking features.
Despite the success of neural models in IR, a major performance bottleneck lies in the availability of large scale, high-quality and labeled datasets as deep neural models often have a large number of parameters to learn \citep{Dehghani2017NeuralRM}.

In recent years, \acp{PTM} have brought a storm and fueled a paradigm shift in \ac{NLP} \citep{qiu2020pre}. 
The idea is to firstly pre-train models with self-supervised language modeling, e.g., predicting the probability of a masked token, and then adapt the pre-trained model to downstream tasks by introducing a small number of additional parameters and fine-tuning them with some task-specific objectives.
As is demonstrated in recent works \citep{peters2018deep, Howard2018UniversalLM}, these pre-trained models are able to capture a decent amount of linguistic knowledge as well as factual knowledge, which are beneficial for downstream tasks and can avoid learning such knowledge from scratch. 
Moreover, with the increasing amount of computational power and the emergence of the Transformer architecture \citep{vaswani2017attention}, we can further improve the capacity of pre-trained models by updating the parameter scale, e.g., from million-level to billion-level (e.g., BERT \citep{devlin2018bert} and GPT-3 \citep{GPT3}) and even trillion-level (e.g., Switch-Transformers \citep{Fedus2021SwitchTS}).
Both of these are desirable properties for modeling the relevance in IR.
On one hand, pre-trained embeddings, which are learned on huge textual corpus with self-supervised modeling objectives, are able to capture intrinsic semantics inside queries and documents. 
On the other hand, large-scale pre-trained models with deeply stacked Transformers have sufficient modeling capacities to learn complicated relevance patterns between queries and documents. 
Owing to these potential benefits, we have witnessed explosive growth of research interest in exploiting \acp{PTM} in IR \citep{onal2018neural, Lin2021PretrainedTF}. 
Note that in this survey, we focus on \acp{PTM} in text retrieval, which is central to IR. 
Readers who are interested in \acp{PTM} in content-based image retrieval or multi-modal retrieval could refer to \citep{DBLP:journals/corr/abs-2012-00641, DBLP:conf/naacl/FeiYL21}. 

Up to now, numerous studies have been devoted to the application of \acp{PTM} in IR. 
In academia, researchers have carried out a variety of innovation and initiative in the usage of \acp{PTM} in IR.
For example, earlier attempts tried to leverage pre-trained word embeddings to promote ranking models, and have achieved some notable results \citep{onal2018neural}.
More recent works proposed to improve existing pre-trained models by either reforming the model architecture \citep{MacAvaney2020EfficientDR, khattab2020colbert, Gao2021CondenserAP} or considering novel pre-training objectives \citep{chang2020pre, Ma2021PROPPW, Ma2021BPROPBP}, which better meet the requirements of IR.
Meanwhile, in industry, Google’s October 2019 blog post\footnote{\url{https://www.blog.google/products/search/search-language-understanding-bert/}} and Bing's November 2019 blog post\footnote{\url{https://azure.microsoft.com/en-us/blog/bing-delivers-its-largest-improvement-in-search-experience-using-azure-gpus/}} both showed that pre-trained ranking models (e.g., BERT-based models) can better understand the query intent and deliver a more useful result in practical search systems.
Besides, looking at the ranking leaderboard\footnote{\url{https://microsoft.github.io/msmarco/\#docranking}} today, we can see that most top-ranked methods are built on \acp{PTM}, just by looking at the names of these submissions.
Considering the increasing number of studies on \acp{PTM} in IR, we believe that it is the right time to survey the current literature, highlight advantages and limitations of existing methods, and gain some insights for future development.

In this survey, we aim to provide a systematic and comprehensive review of works about \acp{PTM} in IR. 
It covers \acp{PTM} published in major conferences (e.g., SIGIR, TheWebConf, ICLR, WSDM, CIKM, AAAI, ACL, and ECIR) and journals (e.g., TOIS, TKDE, TIST, IP\&M, and TACL) in the fields of deep learning, natural language processing, and information retrieval from the year 2016 to 2021.
There exists some previous works discussing related topics. 
For example, both \citet{onal2018neural} and \citet{guo2020deep} reviewed the landscape of neural retrieval models used in three major IR tasks. They also discussed early usage of pre-trained embeddings in neural ranking models, but did not cover every aspect of \acp{PTM} in IR.
\citet{guo2022semantic} reviewed semantic models for the first-stage retrieval, including early semantic retrieval models, neural retrieval models, and retrieval models based on \acp{PTM}.
More recently, \citet{Lin2021PretrainedTF} provided a thorough survey of transformer-based models for IR, which reviews existing literature on the application of pre-trained contextual embedding in text ranking. 
Different from these works, we make a comprehensive overview of \acp{PTM} applied in IR, including the usage of pre-trained word embeddings as well as the application of pre-trained transformers. More specifically, we reviewed the application of \acp{PTM} in different components of an IR system, including the first-stage retrieval component, the re-ranking component, and other components.
We also describe \acp{PTM} specifically designed for IR tasks, as well as resources for pre-training or fine-tuning ranking models. 
In addition to the model discussion, we also introduce some open challenges and suggest potentially research directions for future works.

The structure of this survey is organized as follows. 
We will firstly provide a systematic overview of IR in Section 2.
Following this, we then review works about \acp{PTM} applied in the retrieval component, the re-ranking component, and other components in Sections 3 to 5, respectively. 
In Section 6, we present works in designing novel \acp{PTM} tailored for IR.
We also summarize available large-scale datasets as well as popular benchmark leaderboards in Section 7.
Finally, we conclude this paper in Section 8 and raise some promising directions for future research.

%% file: Sections/2-background.tex

\chapter{Background}
\label{section:background}

In this section, we describe basic concepts and definitions of IR in a hierarchical manner and briefly review \acp{PTM} in IR.
This background overview can help readers gain basic ideas of IR and lead to a better understanding on how \acp{PTM} can be beneficial for IR.
\section{A Hierarchical View of IR}
\begin{figure*}[t]
	\centering
		\includegraphics[scale=0.35]{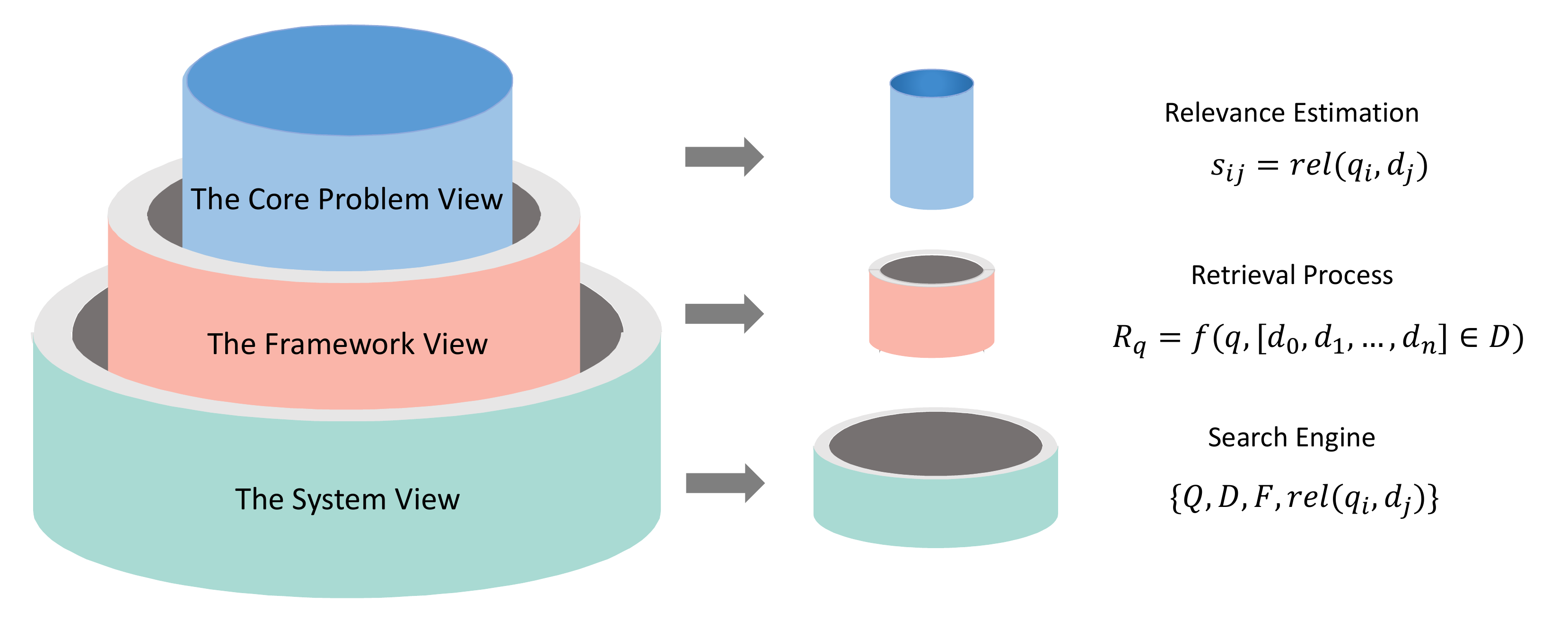}
		  \caption{A Hierarchical View of IR}
  \label{fig:hierarchical_view_of_IR}
\end{figure*}

As is shown in Figure~\ref{fig:hierarchical_view_of_IR}, we illustrate IR by decomposing the search process with a hierarchical view, from the core problem to the framework, to the system.
Specifically, we use capital letters $Q$, $D$, $F$ to denote a set of queries, documents and retrieval functions, and lower-case letters $q$, $d$, $f$ denote a specific instance respectively. 
$rel$ refers to the relevance estimation model which calculate the relevance scores $s_{ij}$ for each ($q_i, d_j$) pair.
$R_q$ denotes returned search results against an issued query $q$.

\subsection{The Core Problem View of IR}
The basic objective of the IR system is to provide relevant information to users in response to their information need. Thus, the most fundamental problem is to estimate the degree of relevance between a query $q$ and a document $d$.
In practice, search begins with the emergence of a user intent which is the main goal a user has when issuing a query into a search engine.
To some extent, the query can be regarded as the representative of the search intent.
Then the mission of the search engine is to return the most ``relevant'' results related to the given query and display these results as a ranked list to the user.
Thus, the better performance of the search engine in terms of estimating the relevance level between $q$ and $d$ the better the user satisfaction.
To evaluate the relevance score of a pair of $q$ and $d$, existing works construct models to consider the correlation between the content of $q$ and $d$ on the basis of different strategies. 
There are three typical groups of these models:
\begin{itemize}
\item \textbf{Classical retrieval models}: 
The key idea of these models is to utilize exact matching signals to design a relevance scoring function.  
Specifically, these models consider easily computed statistics~(e.g., term frequency, document length, and inverse document frequency) of normalized terms matched exactly between $q$ and $d$.
And the sum of contributions from each query term that appears in the document is used to derive the relevance score.
Among these models, BM25~\citep{robertson1995okapi} is shown to be effective and is still regarded as a strong baseline of many retrieval models nowadays. 
Besides BM25 and its variants, there are other representative retrieval functions, such as PIV~\citep{singhal2017pivoted} derived from vector space model, DIR~\citep{zhai2004study} derived using the language modeling approach, PL2~\citep{amati2002probabilistic} based on the divergence from randomness framework, etc. 
However, these models may encounter the ``vocabulary mismatch problem'' due to ``hard'' and exact matching requirements. 
\item \textbf{Learning to Rank (LTR) Models}: 
The key idea of these models is to apply supervised machine learning techniques to solve ranking problems using hand-crafted, manually-engineered features.
Effective features include query-based features~(e.g., query type and query length), document-based features~(e.g., PageRank, document length, number of in-links and number of clicks) and query-document matching features~(e.g., number of occurrences, BM25, N-gram BM25 and edit distance).
According to the number of documents considered in loss functions, LTR models can be grouped into three basic types:
1)~Pointwise approaches which consider individual documents and regard the retrieval problem as classification or regression problem. 
Example models include PRank~(Perceptron Ranking)~\citep{crammer2001pranking} and McRank~\citep{Li2007McRankLT}.
2)~Pairwise approaches which take pairs of documents into consideration. For example, RankNet~\citep{Burges2005LearningTR} is a pairwise method which adopts Cross Entropy as loss function in learning and RankSVM~\citep{herbrich1999support} which performs ranking as a pairwise classification problem and employ the SVM technique to perform the learning task.
3)~Listwise approaches which consider the entire list of documents. For example, LambdaMart~\citep{burges2006learning} trains a ranking function by employing Gradient Descent to minimize a listwise loss function.
Please refer to another survey~\citep{li2014learning} on LTR models for IR for more details.
\item \textbf{Neural Retrieval Models}: 
The key idea of these models is to utilize neural networks to abstract relevance signals for relevance estimation.
These models use the embedding of $q$ and $d$ as the input and are usually trained in an end-to-end manner with relevance labels.
Compared to non-neural models, these models can be trained without handcrafted features.
Without loss of generality, these models can be grouped into representation-focused models, interaction-focused models, and mixed models.
1)~Representation-focused models aims at learning dense vector representations of queries and documents independently.
Then metrics such as cosine similarity and inner products are used to calculate the ``distance'' between queries and documents to estimate the relevance score. 
Example representation-focused models include DSSM~\citep{huang2013learning} and CDSSM~\citep{shen2014latent}, etc.
2)~Interaction-focused models capture ``interactions'' between queries and documents. 
These models utilize a similarity matrix $A$ in which each entry $A_{ij}$ refers to the similarity between embedding of the $i$-th query term and the embedding of the $j$-th document term. 
After constructing the similarity matrix, interaction-based models apply different approaches to extract features that are adopted to produce the query-document relevance score.
Example interaction-focused models include DRMM~\citep{guo2016deep} and convKNRM~\citep{xiong2017end}, etc. 
3)~Mixed models combine the design of the representation-focused component and the interaction-focused component, Duet~\citep{mitra2017learning} and CEDR~\citep{macavaney2019cedr} for example. 
For more detailed information please refer to these earlier surveys~\citep{onal2018neural, guo2020deep} on NeuIR models for IR
\end{itemize}

\subsection{The Framework View of IR}
\begin{figure*}[t]
	\centering
		\includegraphics[scale=0.35]{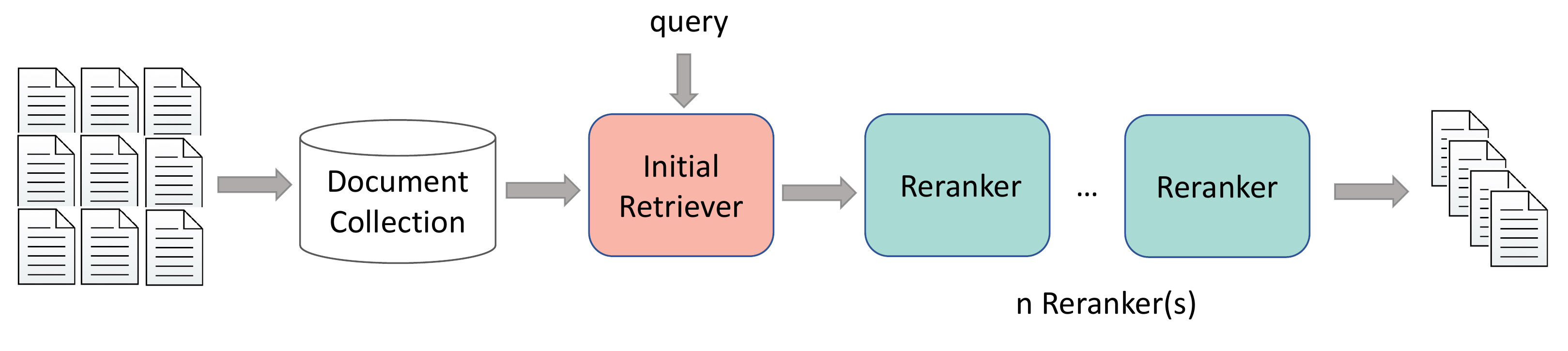}
		  \caption{The retrieval architecture. According to the number of re-rankers, this retrieval process can be defined as Single-stage Retrieval~($n = 0$), Two-stage Retrieval~($n = 1$) and Multi-stage Retrieval~($n \geq 2$).}
  \label{fig:retreival_architecture}
\end{figure*}

Given a document collection $D$, the aim of IR is to provide a search result list $R_q$ where results are ordered in terms of their relevance levels given a query $q$. 
Since the document collection is massive, besides considering effectiveness, a practical IR system needs to give consideration to efficiency as well~\citep{frieder2000efficiency}.
In that regard, in a conventional retrieval architecture, several stages with different focuses on effectiveness and efficiency are built.
We depict a retrieval architecture~($f$ in Figure~\ref{fig:hierarchical_view_of_IR}) in Figure~\ref{fig:retreival_architecture}.
As shown in Figure~\ref{fig:retreival_architecture}, an initial retriever is involved to recall relevant results from a large document collection.
In terms of relevance scores given by the retriever, these initial results are ranked to form an initial result list.
Then this initial result list is passed through $n$ re-rankers to generate the final ranked list which is provided to users.
Each re-ranker receives a ranked list from the previous stage and in turn provides a re-ranked list that contains the same number of or fewer results. 
Although both aiming at estimating relevance levels of query-document pairs, retrievers and re-rankers usually adopt different models.
Since retrievers need to recall relevant documents from a massive document pool, efficiency should be given priority.
In that regard, traditional models such as BM25~\citep{robertson1995okapi} are used to construct initial retrievers. 
As to re-rankers, according to the stage wherein they play a role, re-rankers can be further categorized into early-stage re-rankers and later-stage re-rankers.
Compared to later-stage re-rankers, early-stage re-rankers will focus more on efficiency but will pay more attention to effectiveness than retrievers. 
Since the number of documents considered by later-stage re-rankers is small, later-stage re-rankers will focus more on effectiveness. 
Conventional re-ranking models include learning to rank models~(e.g., RankNet~\citep{Burges2005LearningTR} and LambdaMart~\citep{burges2006learning}) and neural models~(e.g., DRMM~\citep{guo2016deep} and Duet~\citep{mitra2017learning}).

According to the number of re-rankers, the retrieval process can be defined in the following manner~($n$ is the number of re-rankers):
\begin{itemize}
\item \textbf{Single-stage Retrieval~($n=0$)}: the ranked list recalled by the initial retrieval is presented to users without passing through any re-ranker. 
This type of retrieval is applied in early retrieval frameworks such as boolean retrieval and scenarios in which the exact matching is sufficient and preferential.
\item \textbf{Two-stage Retrieval~($n=1$)}: besides the first-stage retrieval, existing IR frameworks also utilize a reranker to further improve the quality of the ranked list.
Features that are not involved in the first-stage retrieval, such as multi-modal features, collected user behaviors and knowledge graphs, are also considered in the re-ranking stage. 
\item \textbf{Multi-stage Retrieval~($n \geq 2$)}: a multi-stage retrieval architecture comprises more than one reranking stage.
Different re-rankers may adopt diverse structures and take advantage of different information sources.
\end{itemize}

\subsection{The System View of IR}

\begin{figure*}[!t]
	\centering
		\includegraphics[scale=0.39]{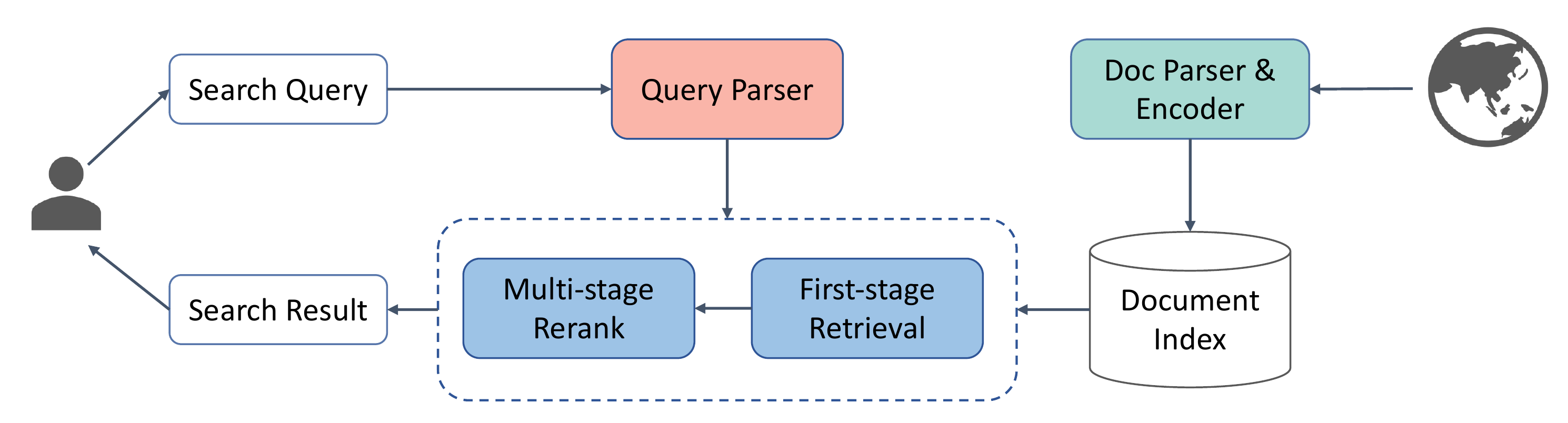}
		  \caption{The framework of a practical search system.}
  \label{fig:search_system}
\end{figure*}

As a practical system, the search system enables end users to perform IR tasks. 
Besides considering effectiveness and efficiency, a good search system should also be user-friendly.
Hence, a good search system needs to deal with different issues existing in the real-world usage which require different components to cooperate.
We depict the conventional framework of a search system in Figure~\ref{fig:search_system}.
The search query issued by a user may be short, ambiguous and sometimes miss-spelt. 
In that regard, a query parser is needed to operate the original query and convert it to a query representation which can reveal the user's true intent to some extent.
The operations on the original query may include rewriting,  expansion and so on.
From the document side, since different web documents have different page structures to organize the content, a document parser/encoder is then essential to process and index web pages. 
A document parser/encoder can also secure the speed in finding relevant documents for a given search query.
Without the document index, the search system would need to scan every document in the corpus, which is time-consuming and requires considerable computing power. 
Besides the query parser and document parser/encoder, the retrieval \& ranking component which is described above is used to provide most relevant results to the user.
In the framework of a search system, the core parts are data structure and storage which are considered in the document component.
Delving into the history of the document index, we observe a paradigm shift from the symbolic search system to the neural search system.
In the following, we briefly introduce how these two systems index documents and also their pros and cons. 

\begin{itemize}
	\item \textbf{Symbolic search system}: 
In a symbolic search system, rules are required to build the document parser which indexes, filters and sorts documents by a variety of criteria, and then translate this data into symbols that the system can understand. 
Hence the name, symbolic search.
Especially, symbolic search system will index documents to build an inverted index which consists of two parts: a dictionary and postings.
The dictionary contains all terms that appear in the document collection.
Then for each term, a list that records which documents the term occurs in is generated.
Each item in the list is called a posting~(or post).
The list is conventionally called a posting list~(or inverted list).
The pros of symbolic search systems are the fast retrieval ability and the provided result is interpretable while the cons are that these systems are stuck using one language and require high maintenance cost~\citep{Manning2005IntroductionTI}.
	\item \textbf{Neural search system}:
While the symbolic search system focuses more on ``exact match'', a neural search system attempts to capture ``semantic match''. 
Instead of designing a set of rules, the neural search system applies pre-trained models to obtain low-dimensional dense representations of documents, which develops a generalized ability of the search system to find relevant results.
The document index in neural search systems is called vector index.
Compared to symbolic search systems, neural search systems are more resilient to noise and easy to extend and scale which are the pros.
The cons of neural search systems include less explainability and the need of lots of data for training~\citep{Mitra2018AnIT}.
\end{itemize}
 
After building the document index~(inverted index or vector index), the search query and documents will be fed into retrieval and re-ranking stages which are elaborated in the above.
In the retrieval and re-ranking stages, symbolic search systems prefer term-based models and learning to rank models, while neural search systems adopt more dense retrieval models and neural ranking models.

\section{A Brief Overview of \acp{PTM} in IR}
Deep learning models are data-hungry.
Especially for models with a massive number of parameters, large datasets are needed to fully learn model parameters and circumvent overfitting issues.
However, building a large-scale labeled dataset for IR is a laborious, expensive and time-taking task.
In contrast, constructing large-scale yet unlabeled corpora~(e.g., crawled web pages and search logs) is much easier.
Thus, an intuitive way is to employ \acp{PTM} to exploit the corpora to learn a better initialization of model parameters. 
Then, the workflow becomes: 1) \acp{PTM} are first applied to learn either good representations of texts or better interaction between text-pairs based on unlabeled datasets; 2) the learned representations/interactions are then fine-tuning and used for downstream tasks.
Specifically, depending on the target downstream task, there exist different options for the fine-tuning: 1) Full fine-tuning: fine-tuning all weights with the data from the downstream task; 2) Partial fine-tuning: fine-tuning partial weights that are specific to the downstream task while freezing the other weights; 3) Freezing the weights: using the representation from the frozen weight to solve the downstream task.
Existing works show that learned representations or interactions extracted from the \acp{PTM} are beneficial for many IR tasks such as document retrieval and re-ranking \citep{guo2016deep, Lin2021PretrainedTF}.
In this Section, we briefly overview typical \acp{PTM} in IR and introduce how they benefit IR in different stages of the search system.
The purpose of this section is to help readers to gain basic knowledge of pre-training methods designed for IR tasks.

The development of \acp{PTM} in IR has roughly gone through two phases.
During the 2010s, in the first phase, word embedding methods have been investigated to develop meaningful representations of words. While recently, in the second phase, transformer-based methods are proposed to gain better representations or interactions of texts by considering more sophisticated model structures and pre-training objectives. We briefly overview these two methods and their relationship to IR.

\subsection{Word Embedding Methods}\label{sec:wd-emb-method}

An embedding refers to a representation of items in a new space where the properties of items and the relationship between these items are preserved.
Then the relatedness of items can be computed based on the notion of similarity in this new space.
In that regard, if the item representations are close to one another means that those items are close to one another.
Word embedding methods learn word representation by setting up an unsupervised prediction task which enables pre-training in a large corpus before using the representation in downstream tasks.
Specifically, the objective is to have words with similar contexts occupy close spatial positions in the new space.
This section briefly overviews classical word embedding methods and their usages in IR tasks.
Classical word embedding methods can be categorized into the following groups:

\begin{itemize}
    \item \textbf{Word2vec}:
In Word2vec approaches~\citep{Mikolov2013EfficientEO, mikolov2013distributed, mikolov2013linguistic}, the word embedding of a term is learned by considering its neighbours within a fixed size window over the text.
There are two architectures, i.e., skip-gram and continuous bag-of-words~(CBOW). 
Both architectures apply a shallow neural model with one hidden states.
For the skip-gram architecture, given a center word, the model learns to predict the most likely words in a fixed-sized window around it. 
For the CBOW architecture, in contrast, the model learns to predict the center word based on the context words.
Since the skip-gram architecture creates more training samples from the same window of text, it trains slower than the CBOW model during training phase~\citep{Mikolov2013EfficientEO}.
    \item \textbf{GloVe}:
\citet{pennington2014glove} proposed GloVe that generates global vectors for word representation. 
Unlike training on individual term-neighbor pairs as in word2vec approaches, GloVe performs training on aggregated global word-word co-occurrence statistics from a corpus.
Different from applying a feedforward neural model, GloVe constructs a word-context matrix, i.e., for each ``word'', how frequently we see this word in some ``context'' can be counted.
Then the matrix factorization technique is utilized to yield a lower-dimensional matrix~(embedding matrix) where each row refers to a vector representation~(word embedding) for a corresponding word.
    \item \textbf{Paragraph2vec}: Paragraph2vec~\citep{le2014distributed}, also known as Doc2vec, is another widely used technique that creates an embedding of a generic block of text, such as sentences, paragraphs and documents.
Expanding upon the Word2vec, Paragraph2vec adds another vector that represents the paragraph ID to the input. 
In that regard, while training the word embedding, the numeric representation of the paragraph can also be obtained.
In the context of IR tasks, \citet{ai2016analysis} and \citet{ai2016improving} proposed a number of changes tailored for IR to the original Paragraph2vec, i.e., document frequency based negative sampling and document length based regularization.

\end{itemize}

Unsupervised and pre-trained word embeddings can be incorporated into IR models and enhance the performance of these models due to their great abilities in capturing semantic and syntactic properties of the input texts.
\textbf{word embeddings are used to refine term weighting schemes in the inverted index}. For example, \citet{zheng2015learning} proposed DeepTR that leverages pre-trained word embeddings learned by the CBOW-based Word2vec.
DeepTR can estimate the term importance and replaces classical term weighting schemes, such as Term Frequency~(TF), in the inverted index so as to improve the retrieval performance.
Moreover, \textbf{word embeddings are applied to better estimate the matching levels of queries and documents}.
For example, \citet{zamani2018neural} proposed SNRM that learns sparse representation for each query and document based on pre-trained word embeddings to better capture semantic relationships between them.
They then constructed an inverted index based on the learned sparse representation which enhances the performance of retrieval.
\citet{gysel2018neural} proposed the Neural Vector Space Model~(NVSM) that is a pre-trained word embeddings method tailored for IR. 
In the NVSM paradigm, they learn low-dimensional representations of words and documents from scratch using gradient descent and rank documents according to their similarity with query representations that are composed of word representations.
Furthermore, \textbf{word embeddings are adopted to benefit crucial IR-related tasks, e.g., query suggestion and document summarization}. 
For example, \citet{dehghani2017learning} used word2vec as an input to encode queries and then feed the query representations into a customized sequence-to-sequence model to deal with the session-based query suggestion problem.
\citet{yin2015optimizing} bulit a CNN-based summarizer, named DivSelect+CNNLM, to enhance the performance of the extractive summarization.
Specifically, the CNNLM module is pre-trained on a large corpus to learn better sentence representations by capturing more internal semantic features.

\subsection{Transformer-based Methods}

Although word embedding methods are demonstrated to be beneficial for IR tasks, they can not deal with the context-dependent nature of words and the issue of polysemous.
This motivates attempt at constructing pre-training methods that can learn context-aware representations of words or interactions between words.
Among them, Transformer~\citep{vaswani2017attention} is a successful instance and has been widely adopted in IR scenarios.
This section briefly overviews typical transformer-based methods, including the structures and pre-training objectives.
We also provide examples of using transformer-based methods in IR tasks.

\citet{vaswani2017attention} proposed transformer, an encoder-decoder architecture that consists of stacked self-attention and point-wise, fully connected layers and supplement modules including positional embeddings, layer normalization and residual connections.
Specifically, in the encoding phase, the transformer first calculates an attention score by comparing a given word with each other word in the input sequence. 
The attention score indicates that how much each of the other words should contribute to the next representation of the given word. 
Transformer then utilizes these attention scores to compute a weighted average of the representations of all the words in the input sequence.
The attention mechanism of the decoding phase is similar to the encoding phase.
The difference is that the attention mechanism in the decoding phase only decodes one representation from left to right at a time and each step of the decoding phase takes into account  results decoded in the previous step.
Due to the parallel modeling capabilities of the self-attention mechanism, transformer is able to train big models with extensive parameters using advanced computing devices.
In that regard, transformer has served as the backbone neural structure for the subsequently derived \acp{PTM}.

GPT~\citep{gpt1} and BERT~\citep{devlin2018bert} are two landmark models of transformer-based pre-training methods.
Among them, GPT uses auto-regressive language modeling as the pre-training objective. 
In particular, the objective is to maximize the conditional probabilities of all the words in the context of their corresponding previous words. 
Hence, GPT is good at generation tasks.
And BERT applies auto-encoding language modeling as the pre-training objective and focus more on language understanding and discriminative tasks.
More specifically, two pre-training objectives word together to optimize the parameters of BERT in the pre-training phase: 1)~Masked language modeling~(MLM): tokens are randomly masked with a special token [MASK] and the objective is to predict words at the masked positions in the context of other words; 2)~Next sentence prediction~(NSP): the objective is to predict whether two sentences are coherent with a binary classifier. 

\begin{figure*}[t]
	\centering
		\includegraphics[scale=0.4]{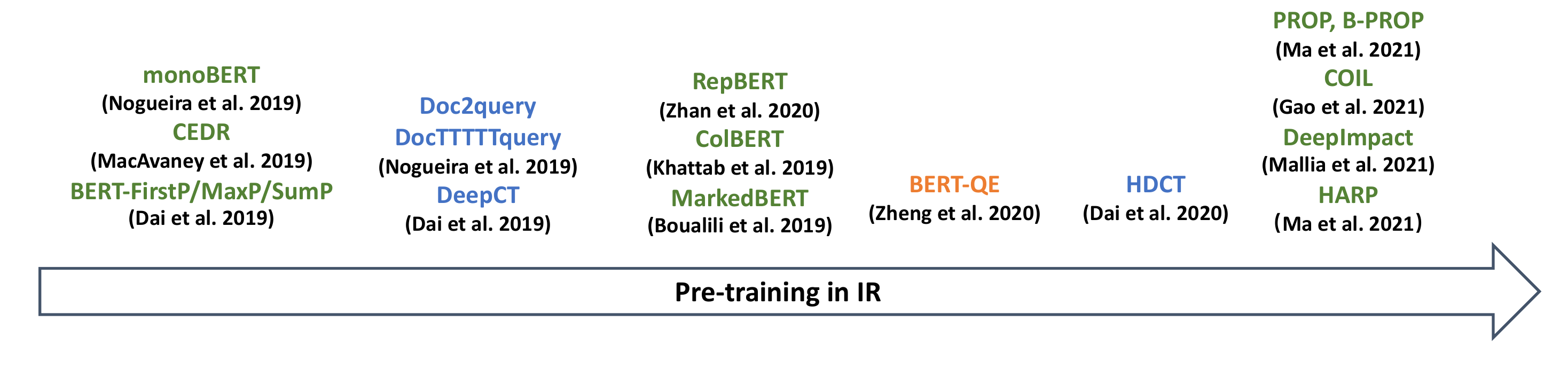}
		  \caption{Recent \acp{PTM} in IR. ``Orange'', ``Green'' and ``Blue'' refer to the ``Query Parser'', ``Retrieval and Rerank'', and ``Doc Parser \& Encoder'' stages for which \acp{PTM} target respectively.}
  \label{fig:PTM_history}
\end{figure*}

Due to their great ability on capturing polysemous disambiguation, syntactic and lexical structures, also the factual knowledge contained in the text, GPT, BERT and their successors have achieved success in IR scenarios.
\textbf{Transformer-based methods are used to estimate the relevance level between the query and the document}.
These \acp{PTM} also have different high-level architectures, such as representation-focused~(e.g., DPR~\citep{karpukhin2020dense}, ColBERT~\citep{khattab2020colbert} and ME-BERT~\citep{luan2020sparse}) and interaction-focused~(e.g., MonoBERT \citep{Nogueira2019PassageRW}, CEDR~\citep{macavaney2019cedr} and duoBERT~\citep{Pradeep2021TheED}).
For example, DPR~(representation-focused) learns dense embeddings for the document with a BERT-based encoder, and queries are encoded with another independent BERT-based encoder.
The outputs of the two encoders are then fed into a ``similarity'' function to obtain the relevance score.
MonoBERT~(interaction-focused) takes the concatenation of the query and document as the input and feeds the [CLS] vector output by BERT to a feed-forward network to obtain the relevance score of the given query and document.
Moreover, \textbf{transformer-based methods also considers the trade-off between efficiency and effectiveness according to the stages~(retrieval or reranking) they targets.}
Especially, for the retrieval stage which focuses more on efficiency, \acp{PTM} are used to improve the performance of retrieval models~(sparse, dense or hybrid).
For example, ColBERT~\citep{khattab2020colbert} generates contextualized term embeddings for queries and documents with a BERT-based dual-encoder and executes two orders-of-magnitude faster per query compared to other baseline models.
In contrast for the re-ranking stage, \acp{PTM} need to deal with a small set of documents and capture more fine-grained relevance signals.
For example, CEDR~\citep{macavaney2019cedr} leverages the contextualized word embeddings of BERT to build a similarity matrix and then feed into an existing interaction-focused neural ranking model such as DRMM and KNRM. The [CLS] vector is also incorporated in CEDR to enhance the model’s signals.
\textbf{Different transformer-based methods are tailored for different components, i.e., ``Query parser'', ``Doc Parser \& Encoder'', and ``Retrieval and Rerank'' in the search system.}
For example, BERT-QE~\citep{zheng2020bert} leverages BERT as the backbone network to expand queries and MeshBART~\citep{chen2020incorporating}  leverages user behavioral patterns such as clicks for generative query suggestion in the ``Query Parser'' component. 
DeepCT~\citep{dai2019context} maps contextualized embeddings learned by BERT to term weights.
Then the predicted term weights are used to replace the original TF field in the inverted index, which refines the ``Doc Parser \& Encoder'' component.
Compared to the ``Query Parser'' and ``Doc Parser \& Encoder'' component, the ``Retrieval and Rerank'' component receives much more attention in the sense that there exist lots of \acp{PTM} designed for this component. 
We show more recent examples in Figure~\ref{fig:PTM_history} where different colors refer to different components on which these \acp{PTM} focus.
Especially, ``Orange'' refers to the ``Query Parser'' component, ``Green'' refers to the ``Retrieval and Rerank'' component and ``Blue'' refers to the ``Doc Parser \& Encoder'' component as shown in Figure~\ref{fig:search_system}.

%% file: Sections/3-retrieval.tex

\chapter{Pre-training Methods Applied in the Retrieval Component} 
\label{section:PTM_in_retrieval}

Traditional search engines rely on term-based retrieval models like BM25~\citep{robertson2009probabilistic} for effective and efficient retrieval. 
Recently, with the rapid progress in representation learning~\citep{bengio2013representation} and pre-training methods~\citep{devlin2018bert, Yang2019XLNetGA, gpt2}, PTMs-based retrieval models have become the popular paradigm to improve retrieval effectiveness. While equipped with \acp{PTM}, retrieval models have achieved great progress in terms of effectiveness~\citep{yan2021unified, karpukhin2020dense}. In this section, we briefly review pre-training methods applied in the retrieval component. Firstly, we give a comprehensive summary of pre-trained retrieval models in terms of model structures. Then, we discuss several challenges and promising topics in terms of the learning of retrieval models.


\section{Basic Model Structure}
From the perspective of representation type and index mode, PTMs-based retrieval models can be devided into three categories~\citep{guo2022semantic}:
1) Sparse Retrieval Models: improve retrieval by obtaining semantic augmented sparse representations and index them with the inverted index for efficient retrieval;
2) Dense Retrieval Models: project input texts (i.e., queries and documents) into standalone dense representations and turn to approximate nearest neighbor search algorithms for fast retrieval;
3) Hybrid Retrieval Models: build sparse and dense retrieval models concurrently to absorb merits of both for better retrieval performance.

\subsection{Sparse Retrieval Models} \label{sec:sparse_retrieval_models}
Sparse retrieval models focus on improving retrieval performance by either enhancing the bag-of-words (BoW) representations in classical term-based methods or mapping input texts into the ``latent word'' space. 
In this framework, queries and documents are represented with high-dimensional sparse embeddings so that the inverted index can be still used for efficient retrieval~\citep{dai2019context, bai2020sparterm}.

With the development of \acp{PTM}, pre-trained models have been widely employed to improve the capacity of sparse retrieval models.
We summarize existing works that apply \acp{PTM} in sparse retrieval models into four classes, including term re-weighting, document expansion, expansion + re-weighting, and sparse representation learning.

\begin{figure*}[t]
	\centering
		\includegraphics[scale=0.3]{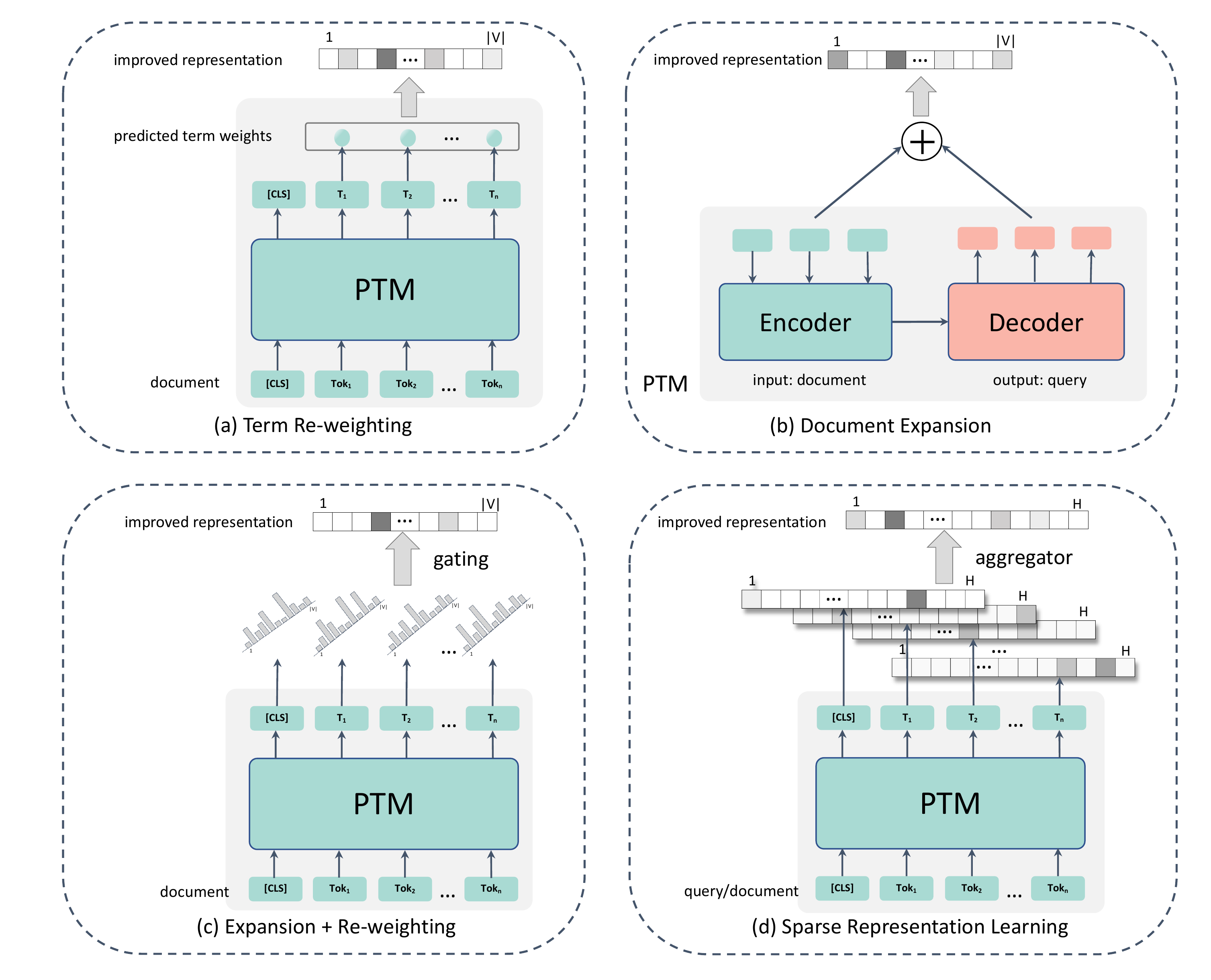}
		  \caption{Four architectures of sparse retrieval models.}
  \label{fig:sparse_retrieval_model}
\end{figure*}

\paragraph{Term Re-weighting} \label{sec:term_reweighting}
One of the most direct ways to improve the term-based retrieval is to measure term weights with contextual semantics, instead of term frequency (TF) (Figure~\ref{fig:sparse_retrieval_model} (a)).
Originally, there have been works utilizing pre-trained word embeddings to estimate term importance. 
Earliest, ~\citet{zheng2015learning} leveraged term weights estimated by pre-trained word embeddings to replace TF in the inverted index to improve the retrieval effectiveness.
Later, ~\citet{frej2020learning} utilized FastText~\citep{bojanowski2017enriching} to estimate the IDF field in the inverted index.
For the above models, the pre-trained word embeddings could be fixed or fine-tuned during the retrieval models training.
Recently, with the development of pre-trained models, there are also explorations to utilize them to estimate term weights. 
For example, ~\citet{dai2020context} used BERT to obtain contextualized token embeddings, and then mapped them to term weights, instead of TF, to build the inverted index.
Later, ~\citet{dai2020context_www} adapted DeepCT~\citep{dai2020context} to estimate term weights for long documents and proposed the HDCT model. It firstly estimates passage-level term weights as the DeepCT does, and then uses a weighted sum to combine them into document-level term weights.

\paragraph{Document Expansion}  \label{sec:document_expansion}
Besides explicitly predicting term weights, augmenting the document with semantically related terms is another practical method (Figure~\ref{fig:sparse_retrieval_model} (b)). 
Based on this, the vocabulary mismatch problem can be alleviated to some extent, and elite terms in the document are promoted at the same time.
In fact, compared with extensive works on query expansion based on PTMs, document expansion are less popular in the IR field. 
Different from early methods that expand documents by mining information from external resources~\citep{sherman2017document, agirre2010document} or the collection itself~\citep{efron2012improving, liu2004cluster, kurland2004corpus}, ~\citet{nogueira2019doc2query} firstly fine-tuned a pre-trained language model T5~\citep{raffel2019exploring} with relevant query-document pairs. The learned model generates multiple queries for each document and appends them to the original document. Then, they used BM25 to retrieve relevant documents based on the expanded document collection.
Later, based on the assumption that document ranking and document expansion tasks share certain inherent relations and can benefit from each other, \citet{yan2021unified} used the document ranking task to enhance the training of document expansion task. They firstly pre-trained the Transformer encoder-decoder architecture~\citep{vaswani2017attention}, where the encoder is pre-trained to support document re-ranking and the decoder is pre-trained for query generation. Then, they conducted a joint fine-tuning process, where a mini-batch is constructed with equal probability from the training data of document ranking or query generation tasks. Finally, the learned Seq2Seq model is used to expand documents as docTTTTTquery~\citep{nogueira2019doc2query} does.

\paragraph{Expansion + Re-weighting}  \label{sec:expansion_reweighting}
Based on the above two methods, a more optimal method is to combine the idea of term re-weighting and document expansion, learning term weights in the whole vocabulary instead of existing tokens in the document (Figure~\ref{fig:sparse_retrieval_model} (c)).
For example, SparTerm~\citep{bai2020sparterm} predicts the term importance distribution in the vocabulary space based on contextual token embeddings got by BERT. Based on this, it re-weights existing and expand terms simultaneously. Moreover, it includes a gating controller to ensure the sparsity of the final representation.
Later, ~\citet{formal2021splade} proposed SPALDE to improve SparTerm~\citep{bai2020sparterm}, which used a saturate function to prevent some terms from dominating the representation and employs a \textit{FLOPS} loss to enable the end-to-end learning.
In addition to doing the expansion and re-weighting simultaneously in a unified framework, ~\citet{mallia2021learning} proposed a simple but effective model called DeepImpact, which leverages docTTTTTquery~\citep{nogueira2019doc2query} to expand documents firstly, and then uses BERT to estimate term importance for appeared terms.

\paragraph{Sparse Representation Learning}  \label{sec:sparse_representation_learning}
Different from the above methods to improve document representations in explicit symbolic space, sparse representation learning methods learn sparse embeddings for queries and documents in the latent word space (Figure~\ref{fig:sparse_retrieval_model} (d)).
SNRM~\citep{zamani2018neural} is the pioneer to learn sparse representations for ad-hoc retrieval. Based on the pre-trained word embeddings, SNRM learns standalone sparse representations for each query and document to capture semantic relationships between them, which shows better retrieval effectiveness over baselines. 
Recently, ~\citet{jang2021uhd} proposed UHD-BERT, which learns extremely high dimensional representations with controllable sparsity based on pre-trained language models. More specifically, it firstly obtains dense token embeddings for queries/documents by BERT and maps them to high-dimensional vectors with a linear layer. Then, the \textit{Winner-Take-All} mechanism is employed to remain top-k dimensions in the dense token embeddings and get the sparse token embeddings. Finally, it generates the sparse query/document representation by token-wise max pooling.
Besides, ~\citet{yamada2021efficient} integrated the learning-to-hash technique into DPR~\citep{karpukhin2020dense} to represent input texts with binary codes. BPR is learned with a multi-task objective, which trains the BERT-based dual-encoder and the hash function in an end-to-end manner. Based on the binary codes of queries and documents, BPR drastically reduces the memory cost of the document index and obtains comparable accuracy on two benchmarks.

\subsection{Dense Retrieval Models}  \label{sec:dense_retrieval_models}
Another research line, namely dense retrieval models, turns to dense representations from sparse representations. 
Dense retrieval models employ the dual-encoder architecture, also known as Siamese network~\citep{bromley1993signature}, to learn low-dimensional dense embeddings for queries and documents. 
Afterward, the learned dense representations are indexed via approximate nearest neighbor (ANN) search algorithms to support online search.

\begin{figure*}[t]
	\centering
		\includegraphics[scale=0.3]{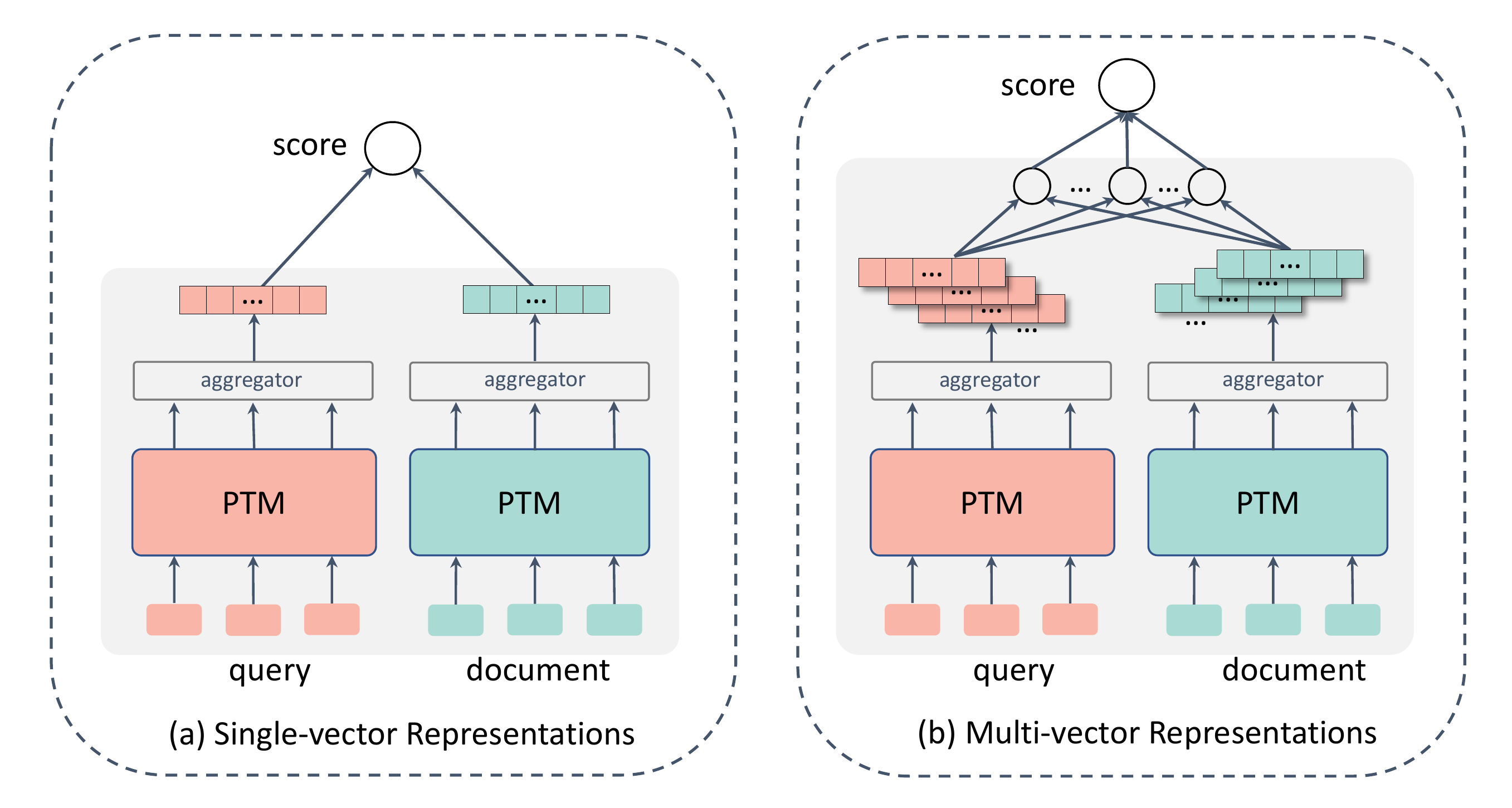}
		  \caption{Basic architectures of dense retrieval models.}
  \label{fig:dense_retrieval_model}
\end{figure*}

Dense retrieval models usually consist of two encoders to learn standalone dense embeddings for queries and documents independently. 
Then, a simple matching function (e.g., dot product or cosine similarity) is used to calculate the relevance scores based on the learned representations.
In this way, the basic architecture of dense retrieval models can be formulated as:
\begin{equation}
\label{eq:rep-focused}
rel(q, d) = f(\phi_{_{PTM}}(q), \varphi_{_{PTM}}(d)),
\end{equation}
where $\phi_{_{PTM}}$ and $\varphi_{_{PTM}}$ are query and document encoders based on pre-training methods, and $f$ is the similarity function.
In the literature, two dense retrieval families have emerged: single-vector representations (Figure~\ref{fig:dense_retrieval_model} (a)), where the entire input text is represented by a single embedding, and multi-vector representations (Figure~\ref{fig:dense_retrieval_model} (b)), where the input text is represented by multiple contextual embeddings.

\paragraph{Single-vector Representation}
Initially, some works used simple heuristic functions to aggregate pre-trained word embeddings and obtained dense representations for queries and documents.
For example, ~\citet{clinchant2013aggregating} presented a document representation model based on pre-trained word embeddings. They used the fisher kernel framework to transform word embeddings into a high-dimensional space and then aggregated them to generate the document representation. 
Afterwards, ~\citet{gillick2018end} obtained query and document representations with the average of pre-trained word embeddings. The surprising experimental results indicate that dense retrieval is a practical alternative to the symbolic-based retrieval models. 
Besieds, ~\citet{gysel2018neural} and ~\citet{agosti2020learning} proposed word-embedding learning methods tailored for IR (see Section~\ref{section:PTM_for_information_retrieval} for details).
However, it is easy to find that obtaining query/document representations by directly aggregating word embeddings would lose contextual semantics and word orders information.
To address this problem, ~\citet{le2014distributed} proposed the Paragraph Vector (PV) algorithm to learn fixed-length representations from variable-length texts. 
Later, ~\citet{ai2016improving} found the unstable performance and limited improvements of PV representations for ad-hoc retrieval and produced modifications to it for IR tasks. 

Except for obtaining dense query/document representations based on pre-trained embeddings, existing attempts at improving the quality of dense retrieval models focuses on finding more powerful representation learning functions. 
This is typically achieved by using a pre-trained language model as the encoder.
One of the representatives that apply pre-trained models for dense retrieval is DPR~\citep{karpukhin2020dense}, which is proposed for OpenQA tasks. DPR learns dense embeddings for queries and passages with two independent BERT-based encoder. Then, relevance scores are calculated with the inner product operation between query and document representations. The results on several OpenQA datasets show that DPR outperforms BM25 and is beneficial for the downstream QA performance.
For ad-hoc retrieval tasks, \citet{DBLP:journals/corr/abs-2006-15498} proposed RepBERT to replace BM25 for the retrieval component. The model architecture of RepBERT is similar to DPR~\citep{karpukhin2020dense} except that RepBERT uses a shared BERT-based encoder for queries and documents.
Similarly, the PTMs-based dense retrieval method also improves conversational search. 
For example, ~\citet{yu2021few} presented ConvDR to learn contextualized BERT embeddings for multi-turn conversational queries and documents respectively, and then retrieves relevant documents using dot products.
Another approach to building a strong dense retriever is to distill the learned knowledge from a more complex model~\citep{tahami2020distilling, lin2020distilling, choi2021improving, hofstatter2020improving}.
For example, ~\citet{tahami2020distilling} utilized the knowledge distillation (KD) technique to distillate the BERT-based cross-encoder network to the dual-encoder model, which heavily increases the retrieval effectiveness.

\paragraph{Multi-vector Representation}
Besides learning a single global representation for queries and documents, another approach is to obtain multiple vectors for them.
A natural method is to take pre-trained word embeddings as term-level representations for queries and documents.
Earliest, ~\citet{Kenter2015Short} proposed to rely only on pre-trained word embeddings for short texts retrieval. They took the cosine similarity between the query word embedding  document word embedding to replace the TF field in BM25 for retrieval, which shows better performance than baselines.
Later, ~\citet{mitra2016dual} proposed to retain dual word embedding spaces. Based on the learned pre-trained word2vec embedding model, query words are mapped into the input space and document words are mapped into the output space. The final relevance score is calculated with aggregated cosine similarities between all query-document word pairs. 

Except for the pre-trained word embeddings, there are also a number of works that employ pre-trained models to learn query/document representations for IR. 
ColBERT~\citep{khattab2020colbert} generates contextualized term embeddings for queries and documents with a BERT-based dual-encoder, and then employs the MaxSim operator to obtain the matching score.
Later, ~\citet{gao2021coil} proposed a similar method, but only calculating similarities between exactly matched terms for queries and documents in the MaxSim operator.
Besides, an alternative way is to employ different encoders for queries and documents based on the heterogeneity between documents and queries. 
For example, ~\citet{luan2020sparse} proposed ME-BERT, which takes the contextualized embedding of CLS as the single-vector query representation and the first $m$ contextualized token embeddings as the multi-vector document representation. Finally, the largest inner product between each document vector with the query vector is token as the relevance score.
Recently, ~\citet{tang2021improving} proposed a novel multi-vector representation method, which clusters BERT-based document term embeddings with k-means to generate multiple representations for each document. Experimental results show that the model can improve retrieval results significantly on several QA datasets .

\subsection{Hybrid Retrieval Models}  \label{sec:hybrid_retrieval_models}
\begin{figure*}[t]
	\centering
		\includegraphics[scale=0.36]{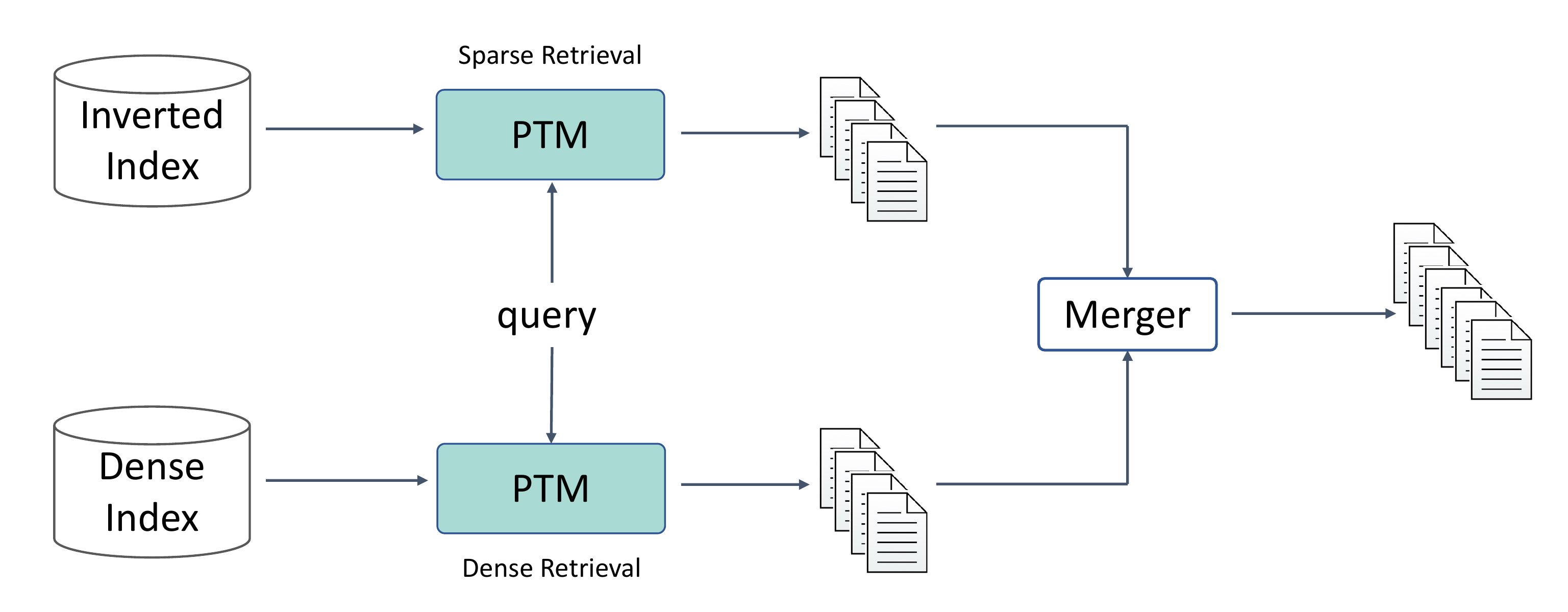}
		  \caption{The architecture of hybrid retrieval models.}
  \label{fig:hybrid_retrieval_model}
\end{figure*}

Sparse retrieval models take a (latent) word as the unit of representations, which can calculate the matching score based on exact matching signals.
On the other hand, dense retrieval methods learn dense embeddings for queries and documents and the relevance is evaluated with soft matching signals. 
To benefit from both of them, hybrid retrieval models learn sparse and dense representations for queries and documents simultaneously and calculate the final relevance scores with a merging method (Figure~\ref{fig:hybrid_retrieval_model}).

To begin with, there are a number of works proposing to combine pre-trained word embeddings with term-based models for the retrieval component.
For example, ~\citet{vulic2015monolingual} combined word embeddings with the language model for monolingual and bilingual retrieval and obtained better results.
Besides, ~\citet{roy2016representing} also proposed to inject pre-trained word embeddings into the standard query likelihood model (QL) for document retrieval.
However, most of these works got the conclusion that only relying on pre-trained word embeddings to build the retrieval model always shows poor performance, unless combining it with the term-based retrieval method. 

With the boosting development of pre-trained models, they are naturally combined with term-based models to enhance retrieval effectiveness. 
~\citet{seo2019real} proposed to learn dense and sparse representations for each phrase in the collection concurrently for OpenQA tasks, where the dense vector is constructed by BERT-based embeddings, and the sparse embedding is the tf-idf representation of the phrase.
Afterwards, ~\citet{lee2019contextualized} proposed to replace the TF-based sparse representation in DenSPI~\citep{seo2019real} with a learned contextual sparse representation based on BERT.
A more simple and direct way to build a hybrid retrieval model is to linearly combine matching scores of a sparse retrieval system and a dense retrieval system using a single trainable weight~\citep{lin2021few, luan2020sparse}. 
For example, ~\citet{luan2020sparse} proposed to linearly combine BM25 and ME-BERT~\citep{luan2020sparse} to produce strong performance.
There are also works using more sophisticated merging methods.
For example, ~\citet{kuzi2020leveraging} leveraged a hybrid approach (BM25 + DE-BERT) with RM3~\citep{abdul2004umass} as the merger for document retrieval, and ~\citet{gao2020complementing} proposed CLEAR to learn the BERT-based dense retriever with the residual of a sparse retrieval model (BM25).

\section{Advanced Topics}  \label{sec:advanced_topics}
Along with the development and achievement of PTMs-based retrieval models, researchers begin to explore more challenging but promising topics.

\subsection{Negative Sampling Strategy}
The negative sampling strategy is a key factor for determining the performance of learned retrieval models.
Generally, hard negative examples are considered as informative negatives, because they can improve the ability of the model to differentiate similar examples.
Thus, how to integrate hard negatives into the learning of PTMs-based retrievers is a widely concerned topic.

One of exemplary methods is the ANCE training method~\citep{xiong2020approximate}, which firstly warms up the RoBERTa-based~\citep{liu2019roberta} dense retriever with BM25 negatives, and then continues the dual-encoder training with the periodically refreshed ANN index for hard negative sampling. Experimental results indicate that ANCE elevates dense retrievers and convincingly surpasses baselines on several benchmarks.
Later, ~\citet{zhan2020learning, zhan2021optimizing} proposed a novel technique for dense retriever training, which constructs the document index based on a warmed-up dense retriever (e.g., ANCE~\citep{xiong2020approximate} or STAR~\citep{zhan2021optimizing}). Then, at each training step, they performed full retrieval based on the fixed document index and updated the query encoder with top retrieved documents as negatives. Experimental results on both passage ranking and document ranking tasks show that the proposed method significantly outperforms all competitive sparse and dense retrieval models.
Recently, ~\citet{hofstatter2021efficiently} argued that previous methods select the training batch with random queries, making in-batch negatives with little information for dense retrievers training. Based on this observation,  they proposed to train dense retrievers with TAS-Balanced batches, which composes training batches with topic-aware query sampling and margin-balanced negative sampling.

\subsection{Joint Learning with Other Components}
To improve retrieval performance, PTMs-based retrieval models can be learned jointly with the index module.
Besides, for different applications, the retrieval component can be learned with downstream components end-to-end, e.g., re-rankers for ad-hoc retrieval and readers for OpenQA.

\paragraph{Joint Learning with Index}
As mentioned above, efficiency is one of the core considerations for the retrieval component. 
To support rapid online search, retrieval systems usually build an index for all documents in the collection.
Specially, for dense retrieval methods described in Section~\ref{sec:dense_retrieval_models}, they usually rely on ANN search algorithms~\citep{aumuller2017ann, echihabi2020return, li2019approximate} to perform efficient retrieval. 
Existing works always separate the dual-encoder learning and ANN index building~\citep{khattab2020colbert, zhan2021optimizing}, which suffer from degraded retrieval performance. 
To address the problem, ~\citet{zhang2021joint} explored the joint training of the dual-encoder and the Product Quantization (PQ)~\citep{jegou2010product} index. They introduced a trainable indexing layer, which is composed of space rotation, coarse quantization and product quantization operations.
Later, ~\citet{zhan2021jointly} proposed JPQ, which firstly utilizes K-Means to generate fixed discrete codes for documents and then only trains the query encoder and PQ Centroid Embeddings jointly. However, this method suffers from a degree of performance loss. 
Further, ~\citet{zhan2021learning} proposed RepCONC, which is capacity to optimize index assignments of document embeddings with a constrained clustering process. Experimental results show the RepCONC achieves better retrieval effectiveness on two benchmarks.

\paragraph{Joint Learning with Re-ranker}
On the basis of the pipeline architecture, most existing works in the IR field focus only on one of components, independently of all the others.
However, separating each component for IR systems building suffers from a few drawbacks and produces sub-optimal performance.
In fact, apart from separately training each component (e.g., retrieval and re-ranking), it has shown that the retrieval and ranking tasks are related with each other~\citep{huang2020embedding, Gao2020ModularizedTR, khattab2020colbert}.
Based on these observations, ~\citet{ren2021rocketqav2} proposed a joint training method for dense retrieval and re-ranking, where the relevance information can be transferred between the two components with a unified list-wise training approach.
Different from this work, ~\citet{zhang2021adversarial} considered to jointly train the two components within an adversarial retriever-ranker (AR2) framework. Within the framework, the retriever aims to recall hard negatives to confuse the re-ranker, and the re-ranker learns to differentiate positives and hard negatives. In this way, the retriever and re-ranker can be enhanced iteratively.

\paragraph{Joint Learning with Reader}
Some studies set about the end-to-end learning of dense retrievers and downstream tasks (e.g., machine reading comprehension (MRC)).
For example, RAG~\citep{lewis2020retrieval} combines a pre-trained dual-encoder (DPR~\citep{karpukhin2020dense}) as the retriever with a pre-trained Seq2Seq model (BART~\citep{lewis2019bart}) as the generator for OpenQA tasks. The query encoder and the generator are fine-tuned end-to-end with the fixed document encoder. The model evaluation on three OpenQA tasks demonstrates the state-of-the-art performance.
Recently, ~\citet{sachan2021end} presented an end-to-end training method for retrieval-augmented OpenQA systems. They built the EMDR$^2$ model, which initializes the dual-encoder retriever with BERT and builds the reader on top of T5. Compared with the stage-wise training, their method allows training signals to flow between the reader and the retriever. Experimental results demonstrate that their method achieves new state-of-the-art results on three benchmarks.

\subsection{Generalization Ability}
In many scenarios outside commercial web search, obtaining training labels is difficult and sometimes infeasible due to privacy constraints (e.g, the medical domain).
Thus, the generalization ability of retrieval models is important in real-world scenarios.
However, many PTMs-based retrieval models have been observed diminishing advantages over term-based retrieval models like BM25 in various benchmarks if they are not fine-tuned with adequate labels (i.e., the zero-shot setup).
Specially, ~\citet{thakur2021beir} studied whether the retriever models can generalize to other domains and concluded that the generalization ability of PTMs-based retrieval models is significantly worse than PTMs-based re-ranking models.

Some early works show great improvement under the zero-shot setting for dual encoders by leveraging strong training losses~\citep{hofstatter2021efficiently} or synthetic data generation~\citep{liang2020embedding,ma2020zero,reddy2021towards}.
For example, TAS-B model~\citep{hofstatter2021efficiently} with the training loss function based on knowledge distillation shows strong generalization capacity and better out-of-distribution performances.
~\citet{ma2020zero} proposed a data augmentation approach to leverage existing QA datasets to train a question generation model given the paired document. Then, the model can be applied to target-domain documents and generates queries for them. Then, these synthetic query-document pairs can be used to train a retrieval model.
Recently, ~\citet{ni2021large} challenged the belief in \citet{thakur2021beir} that models with more interactions between queries and documents have better generalization ability.
They explored the generalization ability of dual-encoder models by scaling up the model size while keeping the bottleneck embedding size fixed. Experimental results on the BEIR dataset~\citep{thakur2021beir} show that scaling up the model size brings significant improvement on a variety of retrieval tasks, especially for out-of-domain generalization.
Besides, \citet{xin2021zero} proposed MoDIR to improve the generalization ability of dense retrievers. Concretely, they introduced an auxiliary domain classifier into the dense retriever training to learn domain-invariant representations, where the retrieval model is not only optimized for the retrieval-orient objective, but also trained to confuse the domain classifier. 

\section{Summary}
This chapter presents how pre-training methods are applied in the retrieval component.

Firstly, we review existing works within three basic model structures, including sparse retrieval models, dense retrieval models, and hybrid retrieval models. 
Sparse retrieval models employ pre-training methods to re-weight terms based on semantic features or map queries/documents into a latent word space to enhance term-based retrieval methods. Due to the sparsity of the representation obtained by sparse retrieval models, they can still utilize the existing inverted index for efficient retrieval.
Dense retrieval models employ PTMs-based dual-encoder architecture to learn standalone low-dimensional dense representations for queries and documents, and then use approximate nearest neighbor search algorithms for fast retrieval. Equipped with pre-training methods, these models often show promising results and naturally obtain increasing research interests in this community.
Hybrid retrieval models are composed of sparse retrieval models and dense retrieval models to absorb merits of both. As expected, these hybrid models usually show better retrieval performance, and at the cost, they require much higher retrieval complexity.

Secondly, we discuss several advanced topics of wide concern to researchers in this community, including negative sampling strategy, joint learning with other components, and generalization ability.
For PTMs-based retrieval models, negative sampling is one of the most important elements for efficient and effective model learning. There have been extensive works focusing on exploring various negative mining methods.
Moreover, the application of PTMs in the retrieval component makes the joint learning of other modules (e.g., index) or downstream tasks possible. Currently, there have been some preliminary works for this topic and it would be a promising direction for the future work.
Although these PTMs-based retrival models have shown inspiring results on several popular benchmarks (e.g., MS MARCO and Natural Questions~\citep{kwiatkowski2019natural}), they are observed reduced advantages if are not fine-tuned with abundant task-specific labeled data. With the release of BEIR\citep{thakur2021beir} benchmark, researchers begin to focus on improving the generalization ability of PTMs-based retrieval models. However, it is still in its infancy stage and worthy of further exploration.

%% file: Sections/4-reranking.tex
\chapter{Pre-training Methods Applied in the Re-ranking Component}
\label{section:PTM_in_reraking}

\begin{figure*}[t]
	\centering
		\includegraphics[scale=0.35]{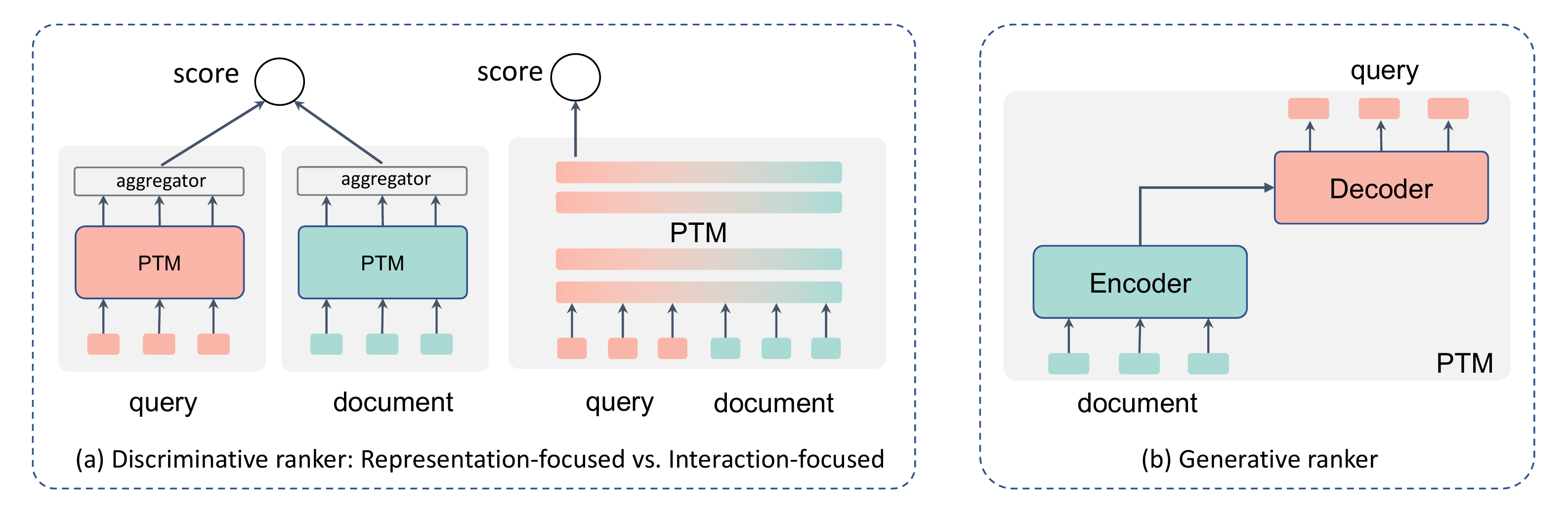}
		  \caption{Two categories of re-ranker.}
  \label{fig:discriminative-generative}
\end{figure*}

In this section, we review previous works applying \acp{PTM} in the re-ranking component.
After the efficient first-stage retriever, there can be a stack of complex re-rankers in the re-ranking stage where the input of each re-ranker comes from the previous one.
Such a multi-stage cascaded architecture is commonly-used both in the industry~\citep{Yin2016RankingRI, Liu2021PretrainedLM, Li2014SemanticMI} and the ranking leaderboard in the academia~\citep{Craswell2021MSMB}.
Generally, \acp{PTM} are often employed to re-rank a small set of candidates provided from the first-stage retriever.
By learning powerful representations or modeling complex interactions between queries and documents, \acp{PTM} have achieved great success compared with previous methods~\citep{Mikolov2013EfficientEO, Lin2021PretrainedTF, Nogueira2019MultiStageDR}.

\section{Basic Model Architecture}

According to the two schools of relevance modeling, i.e., discriminative modeling or generative modeling, in the IR literature~\citep{ponte1998language, robertson2009probabilistic}, the methods applying \acp{PTM} in the re-ranking component can be categorized into three classes:
1) Discriminative Ranking Models: model $P(r,d|q)$ by directly learning a relevance  ``classifier'' from labeled data;
2) Generative Ranking Models: approximate the true relevance distribution $P(r|q,d)$ by modeling the generative process between queries and documents;
3) Hybrid Retrieval Models: joint learn the discriminative model and generative model to leverage merits of both for better ranking performance.

\subsection{Discriminative Ranking Models}

From the very beginning (about 2015-2018), applying \acp{PTM} in the re-ranking component focused on leveraging the pre-trained word embedding such as word2vec \citep{Mikolov2013EfficientEO} and GloVe~\citep{pennington2014glove} into discriminative ranking models~\citep{guo2016deep}.
These word embeddings are mainly used to initialize the embedding layer of ranking models, and other components are usually learned from scratch.
Start with BERT, which pre-trains a Transformer model using self-supervised objectives on large-scale unlabeled corpora, both pre-trained word representations and interactions can be ``transferred'' to the ranking model. 
The former can be used in the same way as previous static word embeddings like word2vec, or like the latter that fine-tunes the whole pre-trained model and only a lightweight task-specific classification layer is learned from scratch.
This is also known as the ``pre-train and fine-tune'' paradigm.
It's more convenient to fine-tune the whole model on downstream tasks as there is no need to design complicated model architectures for each task.
BERT and its successors have achieved great success when applied in the re-ranking component in this way.
This type of \acp{PTM} are generally pre-trained with self-supervised language modeling tasks, and the encoder is employed to build the discriminative ranking model~\citep{devlin2018bert,Yang2019XLNetGA}.
We term this type of \acp{PTM} as discriminative \acp{PTM}.
Following the recipe of NeuIR~\citep{guo2020deep}, there are also two ways in applying \acp{PTM} as the discriminative ranking model on the re-rank component, namely representation-focused models and interaction-focused models. We introduce them in detail in the following.

\begin{figure*}[t]
	\centering
		\includegraphics[scale=0.35]{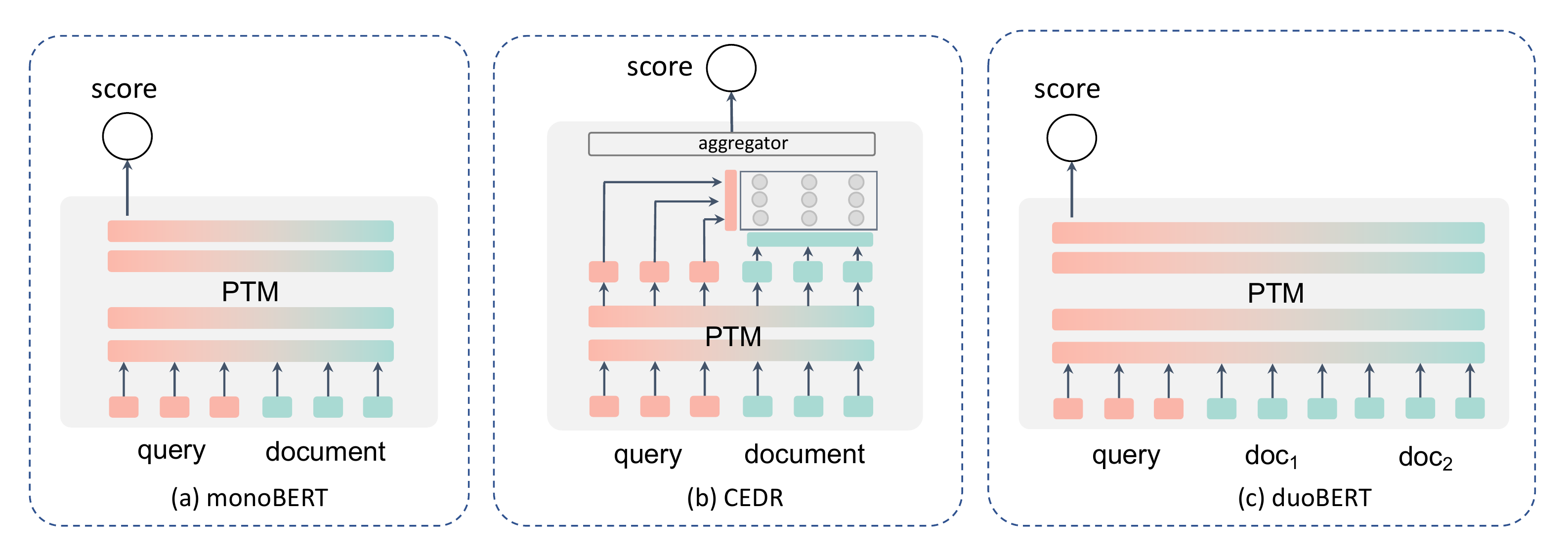}
		  \caption{Three typical interaction-focused discriminative ranking models.}
  \label{fig:three-discriminative-models}
\end{figure*}

\paragraph{Representation-focused Models}
Representation-focused models (Figure \ref{fig:discriminative-generative} (a) left) usually adopt a bi-encoder architecture and encode queries and documents separately, and then the relevance score is computed with simple similarity functions between representations of queries and documents.
Without loss of generality, the representation-focused method could be also abstracted by Eq.~\ref{eq:rep-focused}.
$\phi_{_{PTM}}$ and $\varphi_{_{PTM}}$ are \acp{PTM} which take the raw text of the query or the document as the input, and output one dense representation for each, respectively.
$\phi_{_{PTM}}$ and $\varphi_{_{PTM}}$ could share the parameters or not.
Then, the relevance is computed by simple similarity functions $f$ like cosine or MLP.

In the early days, representation-focused methods often employ pre-trained word embeddings to initialize the representation of input tokens, and the remaining parameters are all randomly initialized.
For example, ARC-I~\citep{Hu2014arc1} trained 50-dimensional word embeddings on Wikipedia and Weibo data using the Word2Vec.
The word embeddings are then fed into convolutional neural networks to obtain text sequence representations, and the relevance score is computed using a MLP based on the two text sequence representations.
They found that fine-tuning the word embedding can further improve the performance compared with fixing them.
More details about the word embedding based ranking model are referred to this survey~\citep{Mitra2018AnIT}.

More recently, the Transformer-based \acp{PTM} are introduced to fine-tune the entire model on downstream tasks, rather than just initializing the word embedding layer.
For example, \citet{Qiao2019UnderstandingTB} proposed to utilize the BERT to encode the query and the document separately, and take the [CLS] embedding of the last layer as their representations and then calculate the ranking score via cosine similarity.
Other studies have shown that using mean pooling on contextual embeddings of the whole input sequence performs better than the [CLS] embedding~\citep{Reimers2019SentenceBERTSE}.
\citet{Qiao2019UnderstandingTB} have shown that representation-based architectures are less effective than interaction-based architectures, but they can be more efficient by utilizing approximate nearest neighbor (ANN) techniques to search from the pre-computed representations.
Thus, the representation-based model architectures are usually applied to the first-stage retrieval phase (see Section \ref{sec:dense_retrieval_models}).

In general, discriminative ranking models can be fine-tuned using the \textit{pointwise}, \textit{pairwise}, or \textit{listwise} learning objectives following the learning to rank literature~\citep{Liu2010LearningTR}.
However, the Transformer-based \acp{PTM} usually limit their input length to 512 due to the quadratic time and memory complexity of self-attention~\citep{devlin2018bert,GPT3}.
Therefore, long documents that contained more than 512 tokens will be truncated before being fed into the model, and more techniques about handling long documents for Transformer-based \acp{PTM} will be introduced in Section~\ref{sec:long-doc-proc}.

\paragraph{Interaction-focused Models}
Interaction-focused models (Figure \ref{fig:discriminative-generative} (a) right) aim to capture low-level interactions between terms in query-document pairs, and then calculate the relevance score based on their interaction features.
For the usage of pre-trained word embeddings in interaction-focused models, they are also used to initialize the representation of input tokens as in representation-focused models~\citep{Mitra2018AnIT}. In this section, we mainly introduce how the Transformer-based \acp{PTM} are used as the interaction-focused model.
Without loss of generality, the interaction-focused method could be abstracted as:
\begin{equation}
\label{equ:interaction}
rel(q, d) = f(\eta_{_{PTM}}(q,d))
\end{equation}
where $\eta_{_{PTM}}$ is the interaction function based on $\acp{PTM}$, and $f$ is the scoring function that estimates the relevance score according to the interaction features.
The input for $\eta_{_{PTM}}$ is a concatenation of the query and the document.
In this way, the interaction of the query and the document could be modeled inside the $\eta_{_{PTM}}$ with the self-attention mechanism.
Note that the interaction cannot be pre-calculated until the query comes, which implies that it's better to use these models re-rank a small set of documents due to the large cost of computing all query-document pairs in the collection.

The most immediate usage of pre-trained Transformers in the interaction-focused model is MonoBERT~\citep{Nogueira2019PassageRW}. It takes the concatenation of the query and the passage as inputs of the BERT, and feeds the [CLS] vector to a feed-forward network to obtain the relevance score.
They take the \textit{pointwise} loss function, i.e., the cross-entropy loss, to fine-tune the BERT model on the MS MARCO passage ranking task~\citep{Craswell2021MSMB}. It is interesting to see that such a direct use of BERT showed outstanding performances compared with previous NeuIR models.
CEDR~\citep{macavaney2019cedr} stacks a traditional neural interaction model upon monoBERT, that is, it leverages the contextualized word embeddings of BERT to build a similarity matrix and then feed into an existing interaction-focused neural ranking model such as DRMM~\citep{guo2016deep} and KNRM~\citep{Xiong2017EndtoEndNA}.
The [CLS] vector is also incorporated in CEDR to enhance the model’s signals.
CEDR is trained using pairwise hinge loss~\citep{Dehghani2017NeuralRM}.
By combining BERT and NeuIR models, CEDR is significantly better than the Vanilla BERT on Robust04 and WebTrack 2012–14.
DuoBERT~\citep{Pradeep2021TheED} takes a sequence comprised of a query and two passages as input and is trained to estimate the positive candidate is more relevant than the negative.
The advantage of DuoBERT is that it can explicitly model the document comparison for pairwise learning objectives.
However, due to the length limitation of the BERT, the whole sequence is truncated to 512 tokens and each passage can have at most 223 tokens.
Though its effectiveness as shown in the passage ranking, the length restrictions largely hinders the application of duoBERT in document ranking tasks.

\subsection{Generative Ranking Models}

\begin{figure*}[t]
	\centering
		\includegraphics[scale=0.35]{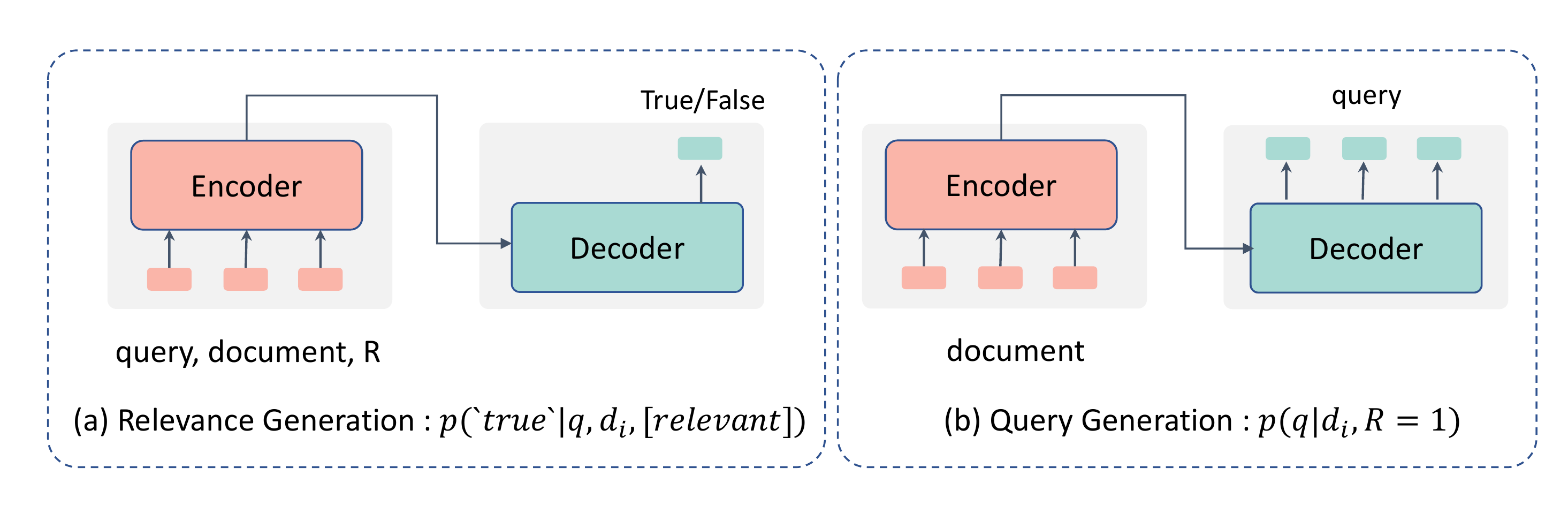}
		  \caption{Two categories of generative ranking models.}
  \label{fig:generative-ranking-models}
\end{figure*}

In addition to discriminative ranking models, researchers have also explored the usage of \acp{PTM} as  generative ranking models~\citep{Roy2016UsingWE,Santos2020BeyondT,Nogueira2020DocumentRW}.
The generative ranking model, which aims to approximate the true relevance distribution, has studied a lot in last century such as statistical language models~\citep{ponte1998language}, and classic probabilistic relevance models~\citep{ robertson1976relevance}.
Statistical language models like the query likelihood model consider the query generation process which ranks documents according to how likely query terms are generated from a document~\citep{ponte1998language, Zhai2008StatisticalLM}.
Classic probabilistic relevance models like Binary Independence Model focuse on describing how a document is generated from a given query~\citep{Lafferty2003ProbabilisticRM, robertson1976relevance}.

For word embedding-based \acp{PTM}, they can be easily incorporated into statistical generative retrieval models to compute the semantic similarity between terms~\citep{Ganguly2015WordEB,Zuccon2015IntegratingAE}.
For example, 
word embeddings can be used to augment the language modeling ~\citep{zamani2016embedding} or the translation modeling~\citep{Zuccon2015IntegratingAE} in the generative ranking model by computing the semantic similarity between terms.  
For Transformer-based \acp{PTM}, they pre-train the decoder of the Transformer or the whole Transformer (i.e., encoder-decoder) with autoregressive language modeling tasks like causal language modeling~\citep{ GPT3}. Then, the pre-trained generative model can be applied to either generate the query or the relevance label.
We term this type of \acp{PTM} as generative \acp{PTM}.
Recent works on applying generative \acp{PTM} to re-ranking are mainly based on the 1) Query Generation process, which is inspired by the query likelihood model.
Another line of researches studied the 2) Relevance Generation process which generates a specified relevance token given the query and the document.

\paragraph{Query Generation}
The first type of generative ranking models is based on the query generation process. 
The basic idea is to rank documents by the likelihood of generating the query from documents using generative \acp{PTM} like GPT~\citep{GPT3} and BART~\citep{lewis2019bart}.
Without loss of generality, the query generative models could be abstracted as
\begin{equation}
\label{equ:query-generation}
rel(q, d) = f(\phi_{_{PTM}}(q|d)) = \prod_{i=1}^{|q|}\phi_{_{PTM}}(q_i|d),
\end{equation}
where $\phi_{_{PTM}}$ is the generative \acp{PTM} and $f$ is a multiplication function $\prod$.
Given the document $d$, each query term $q_i$ is generated one by one and the relevance score is thus obtained by multiplying their normalized probabilities.
The usual approach to train such generative models is to use maximum likelihood estimation (MLE).
Note that, at inference time, the model also uses the Teacher Forcing strategy like the training process.
That is, for each generation, the oracle query term (i.e., ground truth) is used as input for generating the next, instead of model output from a prior time step.


A direct usage of pre-training methods in query generation is to take generative \acp{PTM} like GPT and BART to estimate the probability in generating queries. 
\citet{Santos2020BeyondT} proposed a query generative model for ranking answer passages in QA.
They take the conditional likelihood of generating a question against a passage as the relevance score following Eq.~\ref{equ:query-generation}.
Two types of loss function is proposed to take advantage of both the positive and negative examples: 1) likelihood and unlikelihood loss (LUL) based on MLE; 2) a pairwise ranking loss  (RLL) such as a margin loss based on their likelihood.
Experiments results showed that RLL loss is very helpful for training query generative ranking models.
In addition, they also observed that the generative ranking models can generate fluent questions.
Finally, they found that the query generative models are as effective as simple discriminative ranking models for answer selection.

\paragraph{Relevance Generation} Relevance generation is focused on generating specified relevance tokens by feeding the concatenation of the document and the query into the generative \acp{PTM}, and the probabilities of these relevance labels are treated as relevance scores.
Without loss of generality, the relevance generative models could be reformulated as:
\begin{equation}
\label{equ:relevance-generation}
rel(q, d) = f(\phi_{_{PTM}}(t|q,d)),
\end{equation}
where $t$ is the relevance tokens.
In essence, the relevance generation is a classification task as the model is trained using pointwise loss function on relevance tokens and ranks documents by the probability of predicting the target relevance token.

Considering the relevance token generation is more like a classification task, it can be modeled by both generative \acp{PTM} and discriminative \acp{PTM}.
\citet{Nogueira2020DocumentRW} proposed to use the generative \acp{PTM} T5 for modeling relevance generation.
As T5 is a unified text-to-text language models, they also devised a text-to-text template for the ranking task where the input is ``Query: [q] Document: [d] Relevant:'' and the output is ``true'' or ``false''.
T5 is fine-tuned to generate the target tokens instead of directly producing relevance probabilities.
The probability of the ``true'' token is used to represent the document relevance score, which is normalized with softmax function over the logits of  ``true'' and ``false'' tokens . 
Other target tokens like ``yes/no'' perform worse than the ``true/false'' tokens.
Experiments show T5-3B, which was firstly trained on MS MARCO passage ranking task, outperforms some supervised training models like Birch, BERT-maxP and PARADE, in a zero-shot manner on Robust04.

Moreover, the above method is similar to the prompt learning where the model is guided to predict the ``label'' based on prompts ~\citep{Schick2021ExploitingCF,Schick2021ItsNJ}.
A template and a verbalizer are needed to design first for a given task, where the template is used to transform the original text to a specific form, and the verbalizer is used to project original labels to some words which are fit for the template.
Take a sentiment classification task as an example, assume the template is ``[text] It is [mask]'' in which the token [text] represents the original text, and the token [mask] stands for the verbalized words such as ``great'' and ``terrible''.
These two words are mapped from the positive label and the negative label, respectively.
The \acp{PTM} are trained to predict the probability distribution on the [mask] position given the text with a specific form.
On some NLP tasks, the prompt learning has shown exciting results under the few-shot setting.
It might be that the reformatted task is almost identical to MLM, which makes it a better usage of pre-trained knowledge~\citep{Lester2021ThePO,Li2021PrefixTuningOC}. However, how to leverage the prompt learning to improve the few-shot learning in IR has not been explored at this point.

\subsection{Hybrid models}
Combining the generative and the discriminative modeling leads to the hybrid models.
\citet{Liu2021GeneralizingDR} proposed a multi-task learning approach to jointly learn the discriminative and the generative relevance modeling in a unified pre-trained model.
They assumed that joint these two different types of retrieval modeling leads to better generalized, and hence more effective retrieval model.
To verify this hypothesis, they leveraged the generative \acp{PTM} (i.e., BART) or the discriminative \acp{PTM} (i.e., BERT) to learn discriminative ranking tasks as well as other language generation tasks, such as query generation task, questions generation task, and anchor text generation task.
For the generative \acp{PTM}, they fed the document and the query into the encoder and the decoder respectively.
Then, the query is generated in a sequence-to-sequence manner and the relevance score is calculated by the last token of the entire sequence using a feedforward layer. 
Since the bidirectional attentions in BERT cannot fully adapt to the sequence-to-sequence training strategy, they implemented a mix of attention mechanisms including bidirectional attention, unidirectional attention and cross attention to support sequence-to-sequence tasks.
Their experiments showed that jointly learning discriminative tasks and generative tasks leads to significant improvement on the MS MARCO passage ranking task.

\section{Advanced Topics}
In addition to the direct application of \acp{PTM} in IR, researchers have also developed a considerable amount of studies to address the IR-specific challenges. On one hand, the document length varies significantly across different domains, where \acp{PTM} often fail to address the long document due to the length restriction of the input. On the other hand, \acp{PTM} often consist of a large number of parameters which would increase the search latency. In what follows, we will introduce researches in addressing these two problems.  

\subsection{Long Document Processing Techniques}\label{sec:long-doc-proc}
In the traditional ad-hoc retrieval, documents always contain thousand of tokens in standard TREC datasets~\citep{Voorhees2004Robust,dietz2017trec}.
However, due to the quadratic time and memory complexity of self-attention mechanism in modern Transformer-based~\citep{vaswani2017attention} \acp{PTM}, the length limit of input is always up to 512.
A majority of applications are to segment the long document text into smaller chunks that can be processed by the \acp{PTM} and then do an aggregation over chunks.
Based on the aggregation type, these methods can be broadly categorized into two classes: 1) Passage Score Aggregation: aggregate the relevance score of the query and segmented passage; and 2) Passage Representation Aggregation: aggregate the representations of segmented passages to document representations first and then compute the relevance between query and the aggregated document representations.

\begin{figure*}[t]
	\centering
		\includegraphics[scale=0.3]{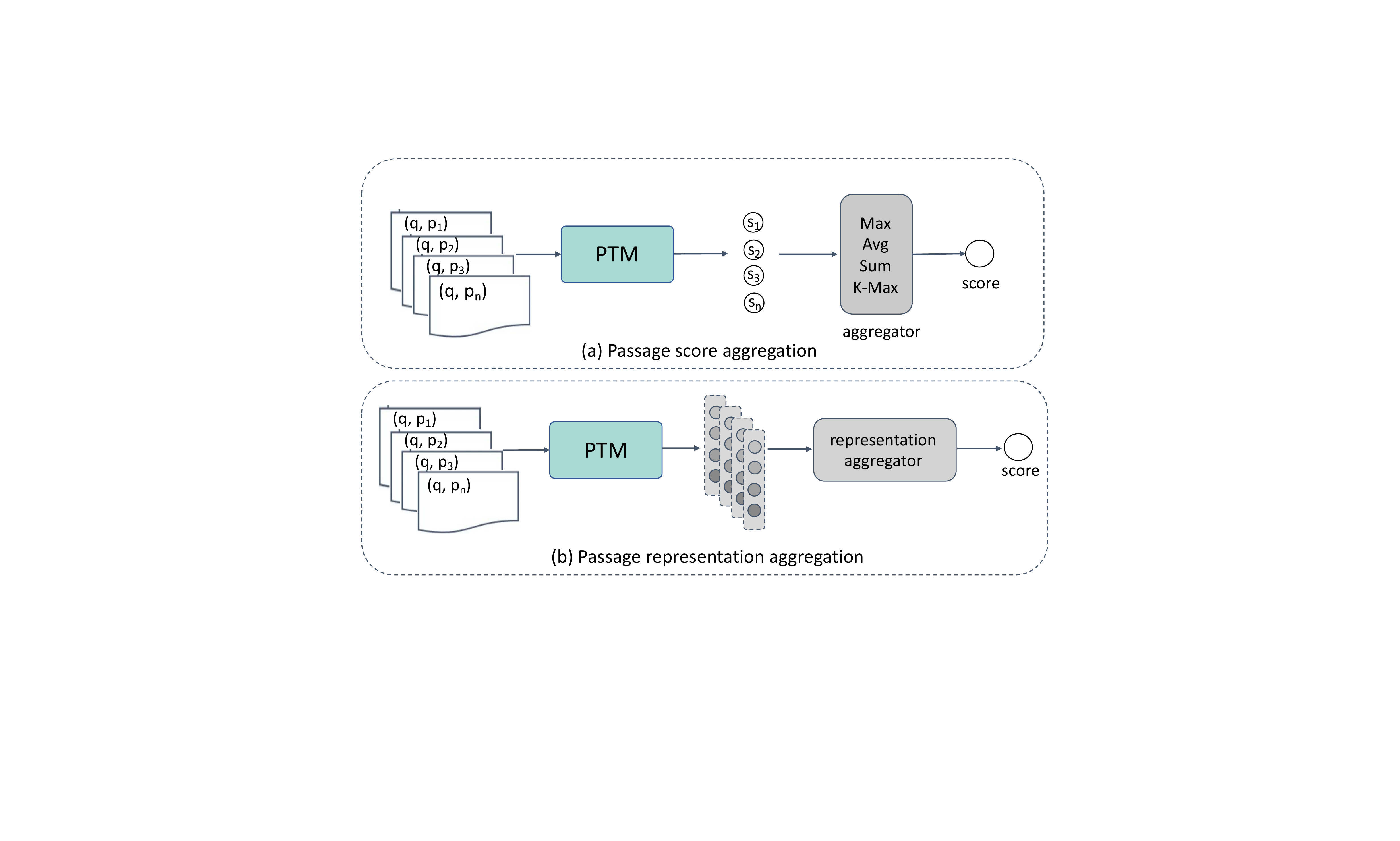}
		  \caption{Two categories of passage aggregation methods.}
  \label{fig:passage-aggregation}
\end{figure*}

\paragraph{Passage Score Aggregation} 
Passage score aggregation is a postprocessing method that only aggregates the relevance score between the query and the segmented passages provided by the \acp{PTM}.
Different methods focus on designing document segmenting and aggregate function~\citep{Dai2019DeeperTU,Hofsttter2021IntraDocumentCL}.

\citet{Dai2019DeeperTU} proposed to split a document into overlapping passages using a 150-word sliding window.
Each passage will be concatenated with the query to input to the BERT, and the relevance of each passage is predicted independently.
Three methods are proposed to aggregate the relevance scores of passages: 1) BERT-firstP that only uses the score of the first passage; 2) BERT-maxP that uses the maximum score of the passages; 3) BERT-sumP that sums all the relevance scores of passages. 
The relevance judgments of segmented passages are consistent with the document, that is, if the document is relevant to a query, all the segmented passages are also relevant to the query and vice versa.
However, according to the \textit{Scope Hypothesis}~\citep{Robertson1994SomeSE}, the document could be partially relevant to a query and thus not all passages are relevant to a query.
There will be noise in the data if we treat all the passages as positive to the query.
BERT-maxP and BERT-sumP perform better than BERT-firstP on traditional ad-hoc retrieval tasks including Robust04 and ClueWeb09-B in their experiments since all passages are taken into account.
But these two methods require more computational cost as all the query-passage pairs need to be trained and predicted while BERT-firstP only considers the first passage of each document.
IDCM~\citep{Hofsttter2021IntraDocumentCL} divides the document into multiple fixed-size windows of 64 words with overlapping of 7 words for both the previous and latter, respectively.
The basic idea is to firstly take a lightweight and fast selection model namely ETM, e.g. Conv-KNRM (CK), to learn to select top-k passages.
And it then takes a slow model like BERT namely ESM to estimate the passage-level relevance score independently and uses a fully-connected network to aggregate the top-k passage score.
Since some operations in the IDCM framework are not differentiable like passage selection module, therefore, they adopts a three-stage optimization pipeline to training the model.
Specifically, ETM and ESM are trained separately where ETM is first optimized on passages and then on full documents by aggregating the score of top passages, ESM is distilled from ETM.
IDCM achieves comparable effectiveness to the BERT-based ranker on two benchmarks including TREC DL 2019 and MS MARCO passage ranking, but with lower computation cost and query latency.
The main restriction is that this method is a little bit complicated on the model framework and the training process.
Although aggregating passage score is simple and effective, it loses the long-range dependence over the whole document as it uses one passage to estimate the relevance independently every time.

\paragraph{Passage Representation Aggregation} 
Instead of only aggregating the passage score, aggregating the passage representation seems more convincing in which the relevance score is estimated by considering all the passages together.

PARADE~\citep{Li2020PARADEPR} segments the long document into a fixed number of overlapping chunks using 225-word sliding windows.
Then all passage representations from a document are aggregated for estimating the document relevance score.
They proposed two types of passage aggregation method: using a mathematical operation such as the elementwise mean, max and sum on the representation vectors; or using a deep neural network including MLP, convolutional neural networks and Transformer layers.
By aggregating the representations with more complicated architectures, PARADE$_{Transformer}$ can significantly improve the performance over passage score aggregation methods like BERT-maxP and other passage representation aggregation methods like PARADE$_{max}$.
PCGM~\citep{Wu2020LeveragingPC} focuses on predicting the sequence of passage-level relevance judgments to avoid splitting a document into independent passages.
It shows the superiority of capturing the context-aware fine-grained passage-level relevance signals.
To be more specific, they first studied the accumulation process of patterns of the passage-level information from a user’s information seeking perspective.
They show the sequence of passage-level cumulative gain can be effectively predicted as a sequence prediction task.
Then, BERT is employed to learn representations of each query-passage pair and then a LSTM network is adopted to aggregate passage representations and predict the passage cumulative gain.
The cumulative gain of the last passage is treated as the document-level gain or the document-level relevance score.
The model is trained on graded passage-level relevance judgments to predict the cumulative gain of previous passages.
Experiments on two Chinese datasets show its effectiveness in improving ranking performance.
The main limitation is that labeling the passage-level relevance judgments is too expensive.

\subsection{Model Acceleration}

Efficiency is one of the major concerns for applying \acp{PTM} in IR as there is always a large-scale data in the real search scenario.
Since the Transformer-based \acp{PTM} often consist of tremendous amount of parameters ranging from millions to billions, this greatly increase the  computational cost and memory storage.
So it's hard to deploy these \acp{PTM} on the online service in real-world applications or on resource-restricted devices considering their requirement of low latency.
To address this issue,
researches have explored several methods to reduce the high computational cost in the re-ranking stage including decoupling the interaction of the query and the document, model distillation, dynamic modeling, and lightweight fine-tuning.

\paragraph{Decouple the Interaction}
One of the bottlenecks that limits the efficiency of the Transformer-based \acp{PTM} comes from the self-attention mechanism.
In the re-ranking stage, the interaction-focused ranking models that apply Transformer-based \acp{PTM} are widely-used and more effective than representation-focused ranking models.
But the representation-focused models are more efficient as they can pre-compute the document representations to reduce the online inference time.
Researchers have studied to incorporate the advantage of the representation-focused architectures into the interaction-focused architectures.

PreTTR~\citep{MacAvaney2020EfficientDR} employed the BERT model and proposed to decouple the low-level interaction of the query and the document via encoding them separately and then interacting in the late BERT layers.
Thus, the document representations can be pre-computed offline and only the query needs to be encoded online.
So most computational budget comes from the interaction of the last few layers now.
When merging the query representation and document representation on layer 11 of BERT, PreTTR achieved a 42X speedup on TREC WebTrack while not significantly reducing the ranking performance.
But merging them at layer 11 performs poorly on the Robust04 dataset.
This indicates that merging the representation of queries and documents at which layer depends on the datasets.
When the query and the document encoding is totally decoupled, it degrades to the representation-focused architecture.
Thus, it's a trade-off between efficiency and effectiveness.
MORES~\citep{Gao2020ModularizedTR} proposed a similar idea to improve the efficiency of the BERT-based re-ranker in which they modularize the Transformer-based neural re-ranker into two separate modules, i.e., text representation module and interaction module.
One of the main differences is that the interaction module in MORES is not a fully cross-attention mechanism.
It canceled the document-to-query attention, and only query-to-document attention is performed followed by query self-attention.
In this way, the document representation is kept unchanged for all queries.
Experiments on MS MARCO passage ranking and TREC 2019 passage ranking showed that 2 layers of lightweight interaction module can achieve ranking performance competitive with a fully interaction-focused architecture while achieving tens or hundred of speedup.

\paragraph{Model Distillation}
Knowledge distillation is a widely used method for reducing the computational cost by transferring knowledge from the teacher to the student.
The basic idea is to learn a smaller model from the outputs of a larger teacher model~\citep{Hinton2015DistillingTK, Sanh2019DistilBERTAD}.
\citet{Hinton2015DistillingTK} proposed a simple yet effective method that transfers the final logits of the teacher on labeled data and unlabeled data to the student where the teacher is first trained on supervised data.
Other studies also investigate to transfer the intermediate hidden states or the attention matrix~\citep{Jiao2020TinyBERTDB}.
For the Transformer-based \acp{PTM}, there are many studies to verify its effectiveness on various tasks including in IR.

\citet{Gao2020UnderstandingBR} investigated three methods to distill BERT for ranking, including only distilling the ranking information of the search task (Ranker Distill), distilling the MLM information over a large text corpus followed by a normal fine-tuning on the search task (LM Distill+Fine-tuning), and distilling both (LM Distill+Ranker Distill).
The teacher model uses BERT-base which contains 12 layers of Transformer, and the student model uses a 4 or 6 layers of Transformer.
Experiments on MS MARCO passage ranking task showed that distilling the ranker behavior alone is not sufficient and LM Distill+Ranker Distill method performs best across all datasets and different size of models.
The 6-layer distilled BERT ranker(2X speedup) using the LM Distill+Fine-tuning method is able to achieve comparable performance to the original BERT, while the performance of the 4-layer distillation BERT ranker (9X speedup) drops significantly.
On top of the TinyBERT model~\citep{Jiao2020TinyBERTDB}, \citet{Chen2021SimplifiedTK} explored to distill the student model with three other kinds of internal weights of the teacher model simultaneously only in the fine-tuning stage, i.e., the attention weight, the hidden state weight, and the embedding weight.
Experiments show that distilling more knowledge from the teacher model can also benefit the ranking.

\paragraph{Dynamic Modeling} Dynamic modeling which can adapt the model structures or parameters to different inputs is another promising method that can improve the efficiency of big models~\citep{Han2021DynamicNN}.
Dynamic modeling can selectively activate some model components of the whole model, such as some layers or a sub-network, conditioned on different inputs, and thus allocate computations on demand at the inference stage.
For example, easy samples will have less computation as they can be predicted quickly with a high confidence.
Early exit is a representative method in this line of research, which allows the examples to exit at early layers of the model without passing through the entire model. 

It is natural to apply the idea of early exit to \acp{PTM} on ranking tasks, since most irrelevant documents can be easily predicted given the query.
\citet{Xin2020EarlyEB} employed such idea from DeeBERT~\citep{Xin2020DeeBERTDE} to the document ranking task.
Specially, extra classification layers are attached to transformer layers of a pre-trained BERT model and then fine-tune the model by simply minimizing the sum of loss functions of all classifiers.
During inference, if the classifier of the $i{th}$ layer is confident about the prediction of the sample, early exiting is performed and subsequent transformer layers are skipped.
Note that the positive confidence threshold and the negative threshold in their paper are set to different values as they assume that positive (relevant) documents need more computations and the confidence score of positive documents is not only the exiting criterion but also the score for re-ranking.
Experiments on the MS MARCO passage ranking dataset showed early exiting is able to accelerate inference by about 2.5X while maintaining the effectiveness of the original model.
Cascade Transformer~\citep{Soldaini2020TheCT} is a sequence of re-rankers built on top of RoBERTa, that is, each re-ranker is a sub-network contained several Transformer layers and a new classification layer, one after another.
When a batch is fed into the Cascade Transformer, each re-ranker will prune a subset of candidates and input the rest to the next until meet the last re-ranker.
In this way, only a small set of candidates in one batch is passed through the whole model and most are pruned early.
To enable this approach, the parameters of all re-rankers are trained in a multi-task learning fashion, in which one of the re-rankers is sampled to train and update the layers  below the selected re-ranker for every mini-batch.
Experiments showed that the Cascade Transformer can get competitive performance to the original RoBERTa while largely reducing the computational cost (over 37\% per batch).

\paragraph{Lightweight Fine-tuning}
The most common way to apply \acp{PTM} is to fine-tune all the parameters given the data from the downstream task.
For the word embeddings, they can be fixed alone or fine-tuned along with the whole neural model without adding too much computation.
However, for the Transformer-based \acp{PTM}, fine-tuning the whole model parameters often requires large computation costs and also storage spaces, especially when serving a large number of tasks with different big models.
With the ever-increasing size of Transformer-based \acp{PTM}, ranging from millions~\citep{devlin2018bert,GPT3} to billions~\citep{GPT3} or even trillions of parameters~\citep{Fedus2021SwitchTS}, fully fine-tuning gradually became impossible for a regular community. 
To mitigate this issue, researchers investigate several lightweight fine-tuning strategies that updates only a small number of extra parameters of \acp{PTM} while keeping most pre-trained parameters frozen.
In this way, we can not only reduce the computation cost to improve the efficiency but also store only one big model and many tunable extra parameters for various tasks.

The intuitive method of lightweight fine-tuning is to freeze some or all pre-trained parameters, but this will hurt the performance greatly without some specific designs~\citep{Houlsby2019ParameterEfficientTL}.
Another line of research studied to insert small neural modules into existing models and only these inserted modules are fine-tuned on the downstream task.
For example, \citet{Houlsby2019ParameterEfficientTL} proposed to insert adapters at each layer, which is a MLP with a non-linear function that projects the input vectors down first and then up.
\citet{Li2021PrefixTuningOC} proposed prefix tuning that prepends several additional prefix tokens to the input or hidden layers, and only these prefix tokens are fine-tuned on downstream tasks.
\citet{Hu2021LoRALA} proposed LoRA that learns low-rank matrices for the attention matrix to approximate parameter updates.
Researchers have also explored these methods into IR tasks.
\citet{Jung2021SemiSiameseBN} examined the above lightweight fine-tuning methods in the \acp{PTM}-based ranking models.
They used a BERT-based bi-encoder architecture for the re-ranking stage.
Experiments on three standard ad-hoc retrieval tasks, including Robust04, ClueWeb09-B and MS MARCO document ranking dataset, showed the effectiveness of these lightweight fine-tuning methods.
In addition, they also proposed a semi-Siamese bi-encoder architecture to reflect the different characteristics of query and document based on the lightweight fine-tuning methods.
For example, when applying prefix-tuning, they add different prefixes for query encoder and document encoder besides a common prefix.
Experiments also demonstrate that such a design can enhance the ranking performance on these datasets.

\section{Summary}

In this chapter, we first review the basic usage of \acp{PTM} when applying in the re-ranking component. 
According to the two schools of relevance modeling in the IR literature, we categorize these works into three classes, i.e., discriminative ranking models, generative ranking models and hybrid ranking models.
1) The word embedding methods are either used to initialize the embedding layer of discriminative neural ranking models or incorporated into the traditional statistical generative models.
But the recent \acp{PTM} pre-train a very deep Transformer model and then fine-tune the whole model on downstream tasks which is proven to be more convenient and powerful.
2) The discriminative ranking models with \acp{PTM} can be modeled with representation-focused architecture or interaction-focused architecture.
The representation-focused architecture is more efficient since it can pre-compute the document representations and only the query is encoded online.
The interaction-focused architecture is more effective but with more computational costs as it needs to encode every query-document pair.
3) The generative ranking models with \acp{PTM} considered two kinds of generation processes, including the query generation and relevance token generation.
The document generation hasn't been studied due to the difficulty of generating long texts conditioned on short texts.
Inspired by the model-based IR system~\citep{metzler2021rethinking}, the model may directly generate the document identifier given the short query instead of the whole document text.
4) The hybrid ranking models jointly learn discriminative ranking objective and query generation using multi-task learning.
Existing approach does not show too much superiority and requires further exploration.
Compared with the generative ranking models, the interaction-focused discriminative ranking models achieved better results on the re-ranking stage~\citep{Santos2020BeyondT}.
But the document identifier generation is also worthy of further exploration considering its efficiency and no needs to store documents.

We then introduced some advanced topics on applying \acp{PTM} in the re-ranking component, such as the long document processing techniques and various strategies to improve its efficiency.
1) Since the quadratic time and memory complexity of self-attention mechanism in the Transformer, most Transformer-based \acp{PTM} limit the input length up to 512 which is often not enough for web documents.
Researchers have studied two approaches to handle long documents including passage score aggregation and passage representation aggregation, and the former is easy to use while the latter performers better~\citep{Li2020PARADEPR}.
2) Although only a small set of documents are re-ranked, efficiency is also one of the major concerns of applying \acp{PTM}, especially the deep Transformer-based \acp{PTM}.
Recent studies mainly focused on decoupling the interaction of the query and the document for the interaction-focused models, model distillation, dynamic modeling and lightweight fine-tuning.
But all existing works have made a compromise, such as increasing training budget (e.g., model distillation and lightweight fine-tuning) or at the expense of performance (e.g., decoupling the interaction and dynamic modeling).
In the future, model quantization an pruning~\citep{Ganesh2021CompressingLT} may be worth trying as they can reduce both the model size and the training cost without losing (too much) performance.

%% file: Sections/5-othercomponent.tex

\chapter{Pre-training Methods Applied in Other Components}
\label{section:PTM_in_other_components}

In this section, we review existing works in applying \acp{PTM} in other components of a search system, such as query expansion, query rewriting, document summarization, snippet generation, etc.
To elaborate, we divide these works into three categories: I) Query Processing, II) User Intent Understanding, and III) Document Summarization.
In the next, we will introduce the pre-training methods applied in these components, respectively.  

\section{Query Processing} \label{sec:query_understanding}
To better bridge the gap between query text and document text, search systems usually contain a query processing module to rephrase the input queries.
Generally, corresponding tasks include query expansion and query rewriting. 

\subsection{Query Expansion} \label{sec:query_expansion}
Query expansion can be considered as an auxiliary task of document ranking, aiming to deal with the vocabulary mismatch problem or to mitigate the gap between queries and documents for better retrieval performance.
Earlier, a large body of work aimed at expanding the original query with the pre-trained word embeddings~\citep{kuzi2016query, Roy2016UsingWE, Diaz2016QueryEW, zamani2016embedding}.
For example, ~\citet{zamani2016embedding} proposed to use word embeddings to incorporate and weight terms that are semantically similar to the query terms and further described two query expansion models which are based on embeddings.
Similarly, ~\citet{kuzi2016query} leveraged the terms to expand the original query or incorporate them with the effective pseudo feedback-based relevance model.

To combine BERT embeddings with probabilistic language models, ~\citet{naseri2021ceqe} developed an unsupervised contextualized query expansion model, namely CEQE, which expands existing queries based on keywords.
Further experiments have demonstrated that CEQE can enhance retrieval effectiveness on multiple standard test collections.
Besides, ~\citet{padaki2020rethinking} proposed that query expansion should be tailored for models like BERT.
Compared to keywords, feeding queries formatted in natural language into BERT-based models may achieve better reranking performance.
In this regard, queries should be expanded with both a rich set of grammar structures and concepts to build word relations.
An intuitive approach is to segment top-ranked documents of a specific query into text chunks and then rank these chunks~\citep{zheng2020bert, zheng2021contextualized}.
For example, ~\citet{zheng2020bert} proposed BERT-QE which leverages BERT as the backbone network to expand queries through three phases: I) rerank candidate documents, II) select relevant text chunks from the top-ranked documents to expand queries, and III) rerank the selected expansion chunks.
These chunks will then be concatenated with the original queries for scoring.

\subsection{Query Rewriting} \label{sec:query_rewriting}

Query rewriting usually aims to 1) map long-tail queries or questions into popular or frequent ones, 2) reformulate ambiguous input queries into well-formed queries to improve retrieval performance.
In the pre-BERT age, some researchers proposed non-contextualized embedding-based approaches for query rewriting~\citep{grbovic2015context, grbovic2016scalable}.
By jointly modeling query content and the corresponding context within a search session, ~\citet{grbovic2015context} propose a novel rewriting method based on a query embedding algorithm.
Their approach maps queries into vectors which are close in the embedding space to allow query expansion via simple K-nearest neighbor search. 

To enhance conversational search, ~\citet{lin2020query} utilized traditional IR query reformulation techniques to realize historical query expansion (HQE) and then applied the T5-base~\citep{raffel2019exploring} model for neural transfer reformulation (NTR), i.e., rewriting a raw utterance into a natural language question without coreference and omission.
There also exists a body of work towards matching user queries or questions to Frequently Asked Questions (FAQs)~\citep{sakata2019faq, mass2020unsupervised, mccreery2020effective}.
For instance, ~\citet{mass2020unsupervised} first employed BERT to calculate the semantic similarity between a query and the candidate FAQs.
They further generated question candidates by fine-tuning GPT-2~\citep{gpt2} in a well-designed unsupervised process and then filtered some noisy candidates according to the semantic similarity.
Besides FAQ retrieval, query rewriting is also applied in spoken language understanding systems for friction reduction~\citep{chen2020pre}, or in dialogue systems to simplify the multi-turn dialogue~\citep{liu2021conversational}.
To reduce the requirement of high-quality query rewriting training pairs, \citet{chen2020pre} proposed a pre-training process which constructs more training objectives by making use of a large amount of readily available historical queries and their Natural Language Understanding (NLU) hypotheses (a serialized word sequence by concatenating domain, intent, slot type and the slot value).

\section{User Intent Understanding}
In complex search scenarios, users may interact with the search system for multiple rounds.
During this process, search systems should understand users' evolving intent to better satisfy their information needs.
Besides modeling users' short-term intent with historical signals, the system can also forwardly provide assistance for search users.
Related tasks include query suggestion, search clarification, and personalized search.

\subsection{Query Suggestion} \label{sec:query_suggestion}
As users' search intents become complex nowadays, a single query usually cannot fulfill their information needs.
In this regard, query/question suggestion techniques provide users with possible future query options, aiming to help users complete their search tasks with less effort in complex search scenarios, e.g., session search or conversational search.
Compared to most previous methods (e.g., HRED-qs~\citep{sordoni2015hierarchical}, ACG~\citep{dehghani2017learning}, and HSCM~\citep{chen2021hybrid}) that used word2vec or GloVe vectors as an input to encode queries, ~\citet{jiang2018rin} constructed a heterogeneous session-flow graph on the AOL dataset and then applied the node2vec~\citep{grover2016node2vec} tool to learn the term embeddings.
The pre-trained term embeddings will then be fed into a reformulation inference network (RIN) to learn a session-level representation.
RIN encodes historical reformulating actions with an RNN-based framework and achieves SOTA performances in both discriminative and generative query suggestion tasks. 

Some other methods have also attempted to employ Transformer-based models for query suggestion~\citep{mustar2020using, chen2020incorporating, mitra2020transformer, rosset2020leading}.
For example, ~\citet{chen2020incorporating} proposed MeshBART which leverages user behavioral pattern such as clicks for generative query suggestion.
To enhance conversational search, ~\citet{rosset2020leading} focused on the usefulness of suggested questions and presented two novel systems.
The first system, namely DeepSuggest, finetunes BERT to rank question candidates by jointly optimizing four learning objectives.
The second one, DeepSuggest-NLG, adopts GPT-2 to generate question suggestions based on the maximum log-likelihood training.
Their approaches leverage the weak supervision signals in the search process, grounding the suggestions to users' information-seeking trajectories and achieving significantly better performance in the usefulness evaluation.
Besides user interactions,~\citet{mitra2020transformer} also utilized search snippet text to recommend related questions in web search.

\subsection{Search Clarification} \label{sec:search_clarification}
As query suggestions are usually presented in a post-search manner, systems can also proactively ask users questions to clarify their information needs and reduce the uncertainty before returning the result list.
Recently, search clarification has attracted much attention in various IR domains such as conversational search and dialogue systems.
To begin with, 
~\citet{habibi2016question} utilized low-dimensional word embeddings learned by word2vec to clarify questions asked by users during a meeting.
From another point, 
 ~\citet{aliannejadi2019asking} proposed BERT-LeaQuR to encode both a query as well as its corresponding candidate questions and then employed a module called NeuQS to select high-quality clarifying questions.
They also presented a new dataset named \textit{Qulac} for conversational search, which collected clarifying questions via crowdsourcing based on the faceted or ambiguous topics in the TREC Web track.
Later,~\citet{hashemi2020guided} introduced Guided Transformer (GT), which utilizes external information such as the top retrieved documents and clarifying questions to learn better representations of input sequences by optimizing a multi-task learning objective.
Extensive experimental results on the \textit{Qulac} dataset suggested that GT substantially outperforms strong baselines in both next clarifying question selection and document retrieval tasks.
Besides, there are also researches focusing on ranking clarifying questions based on natural language inference~\citep{kumar2020ranking} and user engagement prediction~\citep{lotze2021ranking}.
Recently,~\citet{bi2021asking} combined BERT with the maximum-marginal-relevance (MMR) criterion~\citep{carbonell1998use} to clarify user intents with fewer questions as possible.
Their model, namely MMR-BERT, has shown promising efficacy in asking users clarifying questions on the \textit{Qulac} dataset.

\subsection{Personalized Search} \label{sec:personalized_search}
Due to the variety of user propensity, search engines need to provide personalized search services by modeling individual preferences in appropriate scenarios.
A common strategy for personalized search is encoding the search history to capture user's long-term and short-term interests.
Some researchers have attempted to use word embeddings to enhance the personalized search~\citep{kuzi2017query, amer2016toward}.
For example, ~\citet{amer2016toward} realized the personalized query expansion with the word embeddings learned on the user’s profile.
Their work concluded that personalized word embeddings fail to improve the ranking results.
However, ~\citet{kuzi2017query} found that using personalized word embeddings can slightly improve the performance of E-mail search. 

Aware of the remarkable learning power of the Transformer architecture, several recent studies have also focused on building frameworks for personalized search with some Transformer layers~\citep{bi2020transformer, bi2021learning, chen2021hybrid, zhou2020encoding}.
For example,~\citet{zhou2021group} integrated transformer layers with Graph Attention Networks (GANs) and proposed a model named FNPS which considers both search behavior and friend network of users.
To jointly optimize session-level document re-ranking and query suggestion,~\citet{chen2021hybrid} proposed a hybrid framework for session context modeling (HSCM) which leverages both intra-session and cross-session contextual information for personalization.
Unlike general Web search, E-mail search requires personalization in conditions such as recency, user occupation, recipients, and attachments while protecting user privacy.
To this end, ~\citet{bi2021leveraging} leveraged Transformer layers to encode personal e-mail search history, which only contains pre-processed features extracted from raw query and document text.
As different features of one item should be emphasized in various search contexts, a fine-grained review-based transformer model RTM~\citep{bi2021learning} was further proposed to enhance product search by dynamically encoding items at the review level.
Experiment results have indicated both the efficacy of RTM in product search quality and its interpretability.
Most existing personalized approaches do not involve a well-designed pre-training or self-supervised learning (SSL) process, merely utilizing the powerful learning ability of Transformer-like architectures.
Recently, some researchers focused on designing pre-training objectives for personalized search~\citep{zhou2021pssl} or session search~\citep{zhu2021contrastive}.
Their work have shown the great potential of applying contrastive learning in encoding user search history and the content.

\section{Document Summarization} \label{sec:document_summarization}
As most documents contain complicated information, it may take search users a long time to carefully comprehend the whole document.
For users' convenience, modern search engines usually provide a specific piece of text as the preview for a landing page, a.k.a., search snippet.
In some domains, keywords can also be given to enhance the search and classification of the corpus. 

\subsection{Generic Document Summarization} \label{sec:doc_sum}
Generic document summarization aims at automatically compressing given documents into a piece of concise text while keeping salient information.
The task is often generalized into two paradigms: \textit{extractive summarization} and \textit{abstractive summarization}.
In extractive summarization, several sentences are selected from the original document and then concatenated to form a summary, while abstractive methods usually rewrite or paraphrase the document by language generation.
Each paradigm has its own merits and limitations.
For example, extractive summaries are more faithful in content, while they may also have low coherence or consistency between the selected sentences.
Moreover, previous work shows that extractive approaches tend to choose long sentences.
In contrast, abstractive summaries are more flexible while uncontrollable.

Recently, PTMs have been proved effective to be applied in both extractive~\citep{zhang2019hibert, liu2019fine, zhong2020extractive, wang2019self, xu2020unsupervised, zhong2019searching} and abstractive summarization~\citep{zhang2020pegasus, dou2020gsum, lewis2019bart, zou2020pre, saito2020abstractive}.
Earlier,~\citet{yin2015optimizing} built a strong CNN-based summarizer, namely DivSelect+CNNLM, to enhance extractive summarization by projecting sentences into dense distributed representations (\textit{CNNLM}) and then constructing a diversified selection process (\textit{DivSelect}).
The CNNLM module is pre-trained on a large corpus and proved to learn better sentence representations by capturing more internal semantic features.
Their method outperforms many traditional approaches such as LexRank~\citep{erkan2004lexrank} and DivRank~\citep{mei2010divrank} on the DUC 2002/2004 datasets, which can be considered as an early step in adapting PTMs in text summarization.
Besides CNN, pre-trained word embeddings have also been adopted for document summarization~\citep{kobayashi2015summarization, kaageback2014extractive, mohd2020text}.
Generally, they aggregated the word embeddings within a document to represent the whole document and then calculated the semantic similarity at document-level to extract a summary. 

These years have witnessed the superb performance of PTMs such as BERT applied in various NLP tasks.
Document summarization has also been greatly improved with the widespread use of these PTMs.
For instance, ~\citet{zhong2019searching} introduced BERT as external transferable knowledge (contextualized word embeddings) for extractive summarization and reported its superiority compared to word2vec~\citep{Mikolov2013EfficientEO} and GloVe~\citep{pennington2014glove}.
~\citet{zhang2019pretraining} first applied BERT into abstractive summarization via a two-stage decoding process: 1) firstly, generate the draft summary using a left-context-only decoder with copy mechanism; 2) then refine the summary using a refining decoder.
Moreover, \citet{liu2019fine} proposed a general framework called BERTSUM~\footnote{The variants include BERTSUMEXT, BERTSUMABS, and BERTSUMEXTABS (multi-task learning).} for both extractive summarization and abstractive summarization.
Their experiments also indicated that the loss of the extractive task could further improve the abstractive task.
To predict sentences instead of words, HIBERT~\citep{zhang2019hibert} maintains a hierarchical bidirectional transformer architecture and masks documents at sentence-level during encoding.
As most work may cause a mismatch between the the evaluation metrics and the training objective by merely optimizing sentence-level ROUGE, ~\citet{bae2019summary} presented a novel training approach that directly maximizes summary-level ROUGE scores through reinforcement learning (RL).
Their method can achieve better performance in the abstractive summarization task.
To combine auto-encoding with partially auto-regressive language modeling tasks,~\citet{bao2020unilmv2} took Transformer as the backbone network to pre-train a unified language model UniLMv2.
They designed a novel training procedure to jointly pre-train a bidirectional language model (LM) for language understanding and a sequence-to-sequence LM for language generation, namely pseudo-masked language model (PMLM).
Based on this technique, UniLMv2 performs better than other base-size pre-trained models such as BERTSUMABS and MASS in fine-tuning~\citep{song2019mass}.

While most approaches only involve pre-training tasks such as token or sentence masking, BART~\citep{lewis2019bart} corrupts raw text with more noising functions (such as token deletion, sentence permutation, text infilling, and document rotation) and learns a model to reconstruct the original text.
Therefore, BART is particularly effective when fine-tuned for abstractive summarization.
It outperforms the best BERTSUM model by roughly 6.0 points on all ROUGE metrics in both CNN/DailyMail and XSum datasets.
Unlike most previous approaches, MatchSUM bypasses the difficulty of summary-level optimization based on contrastive learning by taking extractive summarization as a semantically text matching problem.
The main point is that a good summary should be more semantically similar to the source document than the other candidates.
Their approach borrows similar ideas from the IR domain and achieves considerable extractive summarization performance on six datasets.
More elaborately, Google proposed a novel framework named PEGASUS~\citep{zhang2020pegasus}, which adopts the gap-sentence generation (GSG) task tailored for abstractive summarization while pre-training.
They hypothesized that exploiting a pre-training objective that is more similar to the downstream task may lead to faster and better performance when fine-tuned.
To this end, gap sentences (indicates the most informational or important sentences within a document) will be selected and used as the target generation text for the remaining content.
As a result, PEGASUS achieves SOTA performance in abstractive summarization on most mainstream public summarization datasets.
Recently, some researchers also focused on I) improving the faithfulness of abstractive summaries by using saliency models or adding some guidances, i.e., CIT~\citep{saito2020abstractive} and GSum~\citep{dou2020gsum}, on II) distilling large pre-trained Transformers for summarization~\citep{shleifer2020pre}, or on III) legal domain related tasks~\citep{huang2021element}.

\subsection{Snippet Generation} \label{sec:snippet_generation}
Different from generic document summarization, search snippets should highlight relevant points in the context of a given query.
Therefore, search snippet generation can be considered as one kind of Query-focused Summarization (QFS).
Similar to generic document summarization, this body of work can also be divided into extractive approaches~\citep{zhu2019transforming, feigenblat2017unsupervised, roitman2020unsupervised} and abstractive approaches~\citep{laskar2020query, baumel2018query, chen2020abstractive, su2020caire, laskar2020wsl}.
As some PTMs are proved to be effective in text generation, most existing work adopted PTMs to generate abstractive snippets.
For instance,~\citet{laskar2020query} proposed a transfer learning technique with Transformer for the Query-Focused Abstractive Summarization (QFAS) task via a two-phase process.
In the first phase, the BERTSUM (mentioned in Sec~$\S$\ref{sec:doc_sum}) model is pre-trained on a generic abstractive summarization corpus.
They further fine-tuned the pre-trained model for the QFAS task on a target domain.
During fine-tuning, they concatenated the query with the document and then fed them into the encoder to incorporate the query relevance.
~\citet{baumel2018query} presented RSA-QFS, which incorporates relevance-aware attention into a pre-trained sequence-to-sequence model~\citep{nema2017diversity} for multi-document summarization.
Despite that modern search engines usually present extractive snippets to search users, less effort has been made in employing PTMs for extractive snippet generation.
One work may be~\citep{zhu2019transforming}, which developed a BERT-based query-focused summarization model.
Based on the model, they constructed massive query-focused summarization examples to enhance the modeling of query relevance and sentence context.
One obstacle in query-focused document summarization may be the lack of proper datasets.
Some attention has also been paid on constructing benchmark datasets of certain scale for this task, e.g., \textit{DUC} 2005-2007 QF-MDS task~\citep{dang2005overview, fisher2006query}, \textit{Debatepedia} (IBM)~\citep{nema2017diversity}, \textit{WikiRef} (Microsoft)~\citep{zhu2019transforming}, \textit{qMDS} (Google)~\citep{kulkarni2020aquamuse}, etc.
Besides retrieval systems, some other approaches~\citep{su2020caire, savery2020question} are more suitable for Question-Answering (QA) system as they combine reading comprehension with language modeling.

\subsection{Keyphrase Extraction} \label{sec:keyphrase_extraction}
Keyphrase extraction or identification aims at extracting a set of informational, topical, and salient phrases from a document.
It can not only provide users a quick view of result documents (similar to document summarization) but may also benefit downstream tasks such as document indexing, document recommendation, and query suggestion.
Most of the existing works formulated keyphrase extraction as a sequential labeling task~\citep{lim2020fine, wu2021unikeyphrase, park2020scientific, sahrawat2020keyphrase, liu2020keyphrase}.
There exists a large body of research aiming at leveraging pre-trained word vectors for keyphrase extraction~\citep{wang2014corpus, qiu2019geoscience, papagiannopoulou2018local, mahata2018key2vec}.
For instance, ~\citet{wang2014corpus} proposed a graph-based ranking approach that uses information supplied by word embedding vectors as the background knowledge.
They further performed keyphrase extraction by constructing a weighted undirected graph for a document to compute the final scores of words. 

From another angle, some work~\citep{sahrawat2020keyphrase, park2020scientific} adopted contextualized embeddings generated by BERT or SciBERT~\citep{beltagy2019scibert} as the input of their BiLSTM-CRF architecture for scientific keyphrase extraction.
~\citet{tang2019progress} used BERT with an attention layer to automatically extract keywords from clinical notes.
From another perspective, ~\citet{sun2020joint} proposed BERT-JointKPE which adopts multi-task learning to chunk self-contained phrases within a document and then rank these phrases by estimating their salience.
Their method inherits the spirit of learning-to-rank approaches and achieves promising keyphrase extraction performance in both the web and scientific domains.

%% file: Sections/6-pretrainingforIR.tex

\chapter{Pre-training Methods Designed for IR}
\label{section:PTM_for_information_retrieval}

In this section, we introduce another line of research on designing \acp{PTM} tailored for IR~\citep{Zamani2017RelevancebasedWE,lee2019latent, chang2020pre, Ma2021PROPPW, ma2021pre, Gao2021CondenserAP, chen2022axiomatically}.
Initially, \acp{PTM} were designed for NLP and the goal is to learn good representations for words or texts.
When applying original \acp{PTM} in IR, studies have demonstrated that they can also benefit many IR tasks, since it's one of the basic requirements for IR to build good representations for queries and documents.
However, the core of IR is to model the notion of \textbf{relevance}~\citep{Lavrenko2017RelevanceBasedLM,Saracevic2016TheNO,Fan2021ALS}, which is not considered in the existing \acp{PTM} designed for NLP.
To address this issue, researchers in the IR community have also started rethinking and exploring new pre-training objectives as well as architectures from the IR perspectives.

Without loss of generality, the general ranking function could be further abstracted as
\begin{equation}\label{algo:abs_ranking}
    rel(q,d) = f(\phi(q),\psi(d), \eta(q,d)),
\end{equation}
where $\phi$ and $\psi$ are representation functions to extract representation features, $\eta$ is the interaction function to extract interaction features, and $f$ is the scoring function which is usually a simple function like cosine or a MLP.
According to the role of the \acp{PTM} in the ranking function, we divide them into two categories: 1) Pre-training Embeddings/Representation Models for IR; 2) Pre-training Interaction Models for IR.

For example, traditional word embedding methods take a single text sequence as input and output a fix-dimensional vector for each word.
So the output word embeddings are usually employed to model the representation functions $\phi,\psi$.
The recent Transformer-based \acp{PTM} have two kinds of pre-training methods based on the input format and the pre-training objectives.
The first one takes a single text sequence as input and learns contextualized word representations with various language modeling tasks~\citep{liu2019roberta,Yang2019XLNetGA}, and this type of \acp{PTM} can be categorized into pre-trained representation models to model $\phi,\psi$.
The other one takes a text sequence pair as input to directly learns their interactions~\citep{devlin2018bert,wang2019structbert,Lan2020ALBERTAL}, and this type of \acp{PTM} can be categorized into pre-trained interaction models.
Note that the pre-trained representation models can also be applied to the  interaction-focused architecture by fine-tuning on labeled data, and vice versa.
However, this will create the pretrain-finetune discrepancy which may not activate their full power of pre-training.

\section{Pre-training Embeddings/Representation Models for IR}\label{sec:6-1}

Pre-trained word embeddings~\citep{mikolov2013distributed,pennington2014glove} are mainly used to initialize the word embedding layer of a neural ranking model while pre-trained representation models~\citep{liu2021gpt,GPT3,liu2019roberta} can be fully ``transferred'' to the IR tasks without designing additional model architectures.
That is, we can fine-tune the entire pre-trained representation models with supervised data on downstream tasks.
Fine-tuning pre-trained models has become the de facto learning paradigm in many fields including NLP and CV.
Methods in this category including word embeddings and the representation models are all pre-trained with self-supervised tasks on large-scale corpora. 
We introduce them next.

\subsection{Static Word Embeddings}

Typical static embedding methods designed for NLP are trained based on word co-occurrence, especially the word proximity, in a large corpus.
By predicting the adjacent word (words) given the context words (word) occurring within a local window, they can capture some lexical, syntactic, and semantic features of words.
Although these word embedding methods have been widely used in neural ranking models and demonstrated to be effective in a number of IR tasks, they are not necessarily equivalent to the primary objective of IR.
The main objective of IR is to predict the words observed in the documents relevant to a particular information need~\citep{Zamani2017RelevancebasedWE}.
Previous studies investigated to design word embedding methods tailor for IR mainly from two aspects:
1) Regularizing the Original Loss towards IR characteristics; 2) Designing New Objectives to capture relevance.
We only briefly describe some representative methods in these two lines of research.

\paragraph{Regularizing the Original Loss}
Some IR-specific characteristics are not considered in the typical word embeddings designed for NLP, such as document-level word frequency and text length, adding these clues to the learning objectives can further improve its effectiveness on IR tasks.
\citet{ai2016improving} found the original paragraph vector (PV)~\citep{le2014distributed} 1) could suppress the importance of frequent words in a document excessively, 2) prone to over-fit short documents during the training process, and 3) ignores to model word-context associations in the learning objective.
Thus, they proposed three modifications to regularize the existing loss function including idf-based negative sampling, introducing L2 to regularize document length, and adding another objective for learning paradigmatic relations.

\paragraph{Designing New Objectives}
Besides regularizing the original loss functions, researchers also explored to design new learning objectives for word embeddings.
\citet{Diaz2016QueryEW} proposed to train local word embeddings in a query-specific manner, that is, using query and the top-k documents retrieved by a statistical language model approach~\citep{Croft2003LanguageMF} to capture the nuances of topic-specific language.
But, this model needs to be trained during the query time and thus is not always practical in real-word applications.
\citet{Zamani2017RelevancebasedWE} pre-trained unsupervised relevance-based word embeddings by predicting the words that occurred in the top-k retrieved documents given the query words under the word2vec framework.
The difference is that they use pseudo-relevance feedback (PRF) models, especially the relevance based language model ~\citep{Lavrenko2017RelevanceBasedLM}, to retrieve documents offline.
They used a very shallow neural network which is a feed-forward neural network with a single linear hidden layer, to train the relevance word embeddings on millions of queries.
Experiments on query expansion task and query classification task showed that the expansion terms chosen by their models are more related to the whole query than word2vec.
\citet{gysel2018neural} proposed another unsupervised model for document retrieval, called NVSM, in which the hypothesis of the optimization objective is that word sequences (i.e., n-grams) extracted from a document should be predictive of that document.
Specifically, multiple phrases of n contiguous words are sampled from a document and then train the averaged word representations of phrases to predict the corresponding document representations.
Experiments show that NVSM outperforms other latent vector space models like word2vec.
Encouraging the n-grams and the document to be close may introduce noise as the randomly sampled n-grams may semantically similar to many documents.

\subsection{Representation Models}

Static word embeddings cannot model polysemy as the use of these words varies across linguistic contexts.
To address this issue, previous methods also proposed to learn context-dependent representations~\citep{Melamud2016context2vecLG,McCann2017LearnedIT,peters2018deep}.
With the development of representation learning, researchers have studied pre-training a whole deep neural model like Transformer~\citep{vaswani2017attention} with self-supervised tasks for the contextualized word representations, and then transferring the entire model to the downstream tasks~\citep{liu2021gpt,GPT3,liu2019roberta}.
The self-supervised tasks are mainly language modeling tasks, such as causal language modeling, masked language modeling and permuted language modeling~\citep{qiu2020pre}.
Although these \acp{PTM} can produce good contextualized word representations, studies have shown that they yield rather bad text sequence embeddings, often worse than averaging GloVe embeddings~\citep{Reimers2019SentenceBERTSE}.
Hence, researchers investigate to pre-train high-quality text sequence representations for queries and documents. 
And the pre-trained representation models are often employed in the representation-focused ranking models.
These works on pre-training representation models for IR are mainly from two aspects: 1) Pre-training Objectives; 2) Model Architectures.

\paragraph{Pre-training Objectives}
According to the underlying hypothesis of learning objectives, previous works can be categorized into two classes.
The first assumes that if the pre-training objective resembles the downstream task, \acp{PTM} can achieve faster and better performance in the fine-tuning stage.
\citet{lee2019latent} proposed a new pre-training task for passage retrieval in open domain question answering (openQA), i.e., Inverse Cloze Task (ICT), where one sentence is randomly sampled from a given passage as pseudo query and the rest sentences are treated as its positive context. 
Inspired by ICT, \citet{chang2020pre} proposed another two tasks to better take advantage of Wikipedia documents.
The first is Body First Selection (BFS) where one sentence from the first section of a Wikipedia page is randomly sampled and another passage from the same page is considered as its positive context.
The other is Wiki Link Prediction (WLP) where the sentence is sampled the same way as in BFS, but the passage is sampled from another hyperlinked Wikipedia page.
These paragraph-level pre-training tasks are pre-trained with a bi-encoder architecture to support the embedding-based dense retrieval.
Experiments on several QA datasets showed that the pre-trained model significantly outperform the widely used BM25 algorithm and the MLM pre-trained models when fine-tuning with a limited number of labeled data.
However, BFS and WLP heavily rely on the special structures of web documents (e.g., multiple paragraph segmentation and hyperlinks), which may hinder their application on a general text corpus.

Another one borrows the idea of information bottleneck theory~\citep{Tishby2015informationbottleneck} which says a good representation is a maximally compressed mapping of the input on the output.
The autoencoder architecture, which performs the compress-then-reconstruct operation to the input, naturally conforms the information bottleneck principle.
Specifically, the general autoencoder consists of an encoder and a decoder, where the encoder maps the input text to representations and the decoder is trained to reconstruct the input text from the representations.
\citet{xiong2021seed} found that the decoder may take shortcuts by exploiting language patterns using its access to previous tokens.
Thus, the vanilla autoencoder is not able to provide high-quality sequence representations.
They proposed SEED which pre-trains autoencoder-based language model with a weak decoder to avoid the bypass effect.
By restricting the model capacity and the attention flexibility of the decoder, the encoder can provide better text representations for dense retrieval.
Experiments on three tasks, including web search, news recommendation, and openQA, demonstrate that SEED is able to boost the effectiveness and few-shot ability significantly.

\paragraph{Model Architectures}\label{sec:p4ir_long}
Due to the quadratic time and memory complexity of self-attention mechanism in vanilla Transformer, the input length of Transformer-based \acp{PTM} is always limited to 512.
However, documents in IR collections are often longer than 512, so vanilla Transformer-based \acp{PTM} are unsuitable to process long documents.
Some studies have investigated designing new architectures to adapt to the IR scenario.
For example, Longformer~\citep{Beltagy2020LongformerTL} proposed to use a combination of a local self-attention and a global attention to sparse the attention matrix.
\citet{Sekulic2020LongformerFM} applied Longformer-based pre-trained models to document ranking.
\citet{Yang2020Beyond5T} proposed Siamese Multi-depth Transformer-based Hierarchical (SMITH) Encoder to handle long document matching tasks.
SMITH learns document representations by hierarchically aggregating the sentence representations from bottom to top.
SMITH is pre-trained with a novel masked sentence block prediction task in addition to MLM task.
Experiments show SMITH outperforms BERT on two document matching tasks by increasing maximum input text length from 512 to 2048.

To learn better text sequence representations, \citet{Gao2021CondenserAP} proposed Condenser which modifies the Transformer architecture by adding a short circuit from the lower layer to the higher layers.
Specifically, for a Transformer model with 12 layers like BERT, they added additional 2 layers on top of the model and also added a short circuit from the 6th layer to the 13th layer.
For the short circuit, the token representations from the 6th layer are directly input to the 13th layer and there is no input from the previous layer, i.e., the 12th layer, except for the special [CLS] token.
They claim that the [CLS] token in the 7-12th layer will focus more on the global meaning of the input text to provide enough information for the top layers to predict the original tokens.
Their experiments showed Condenser improves over standard LM by large margins on various text retrieval and similarity tasks.

\section{Pre-training Interaction Models for IR}
The relevance estimation between a query and a document is to determine whether the information contained in the document satisfy the information need behind the query. Such information could be either a small piece of text span or a long passage, which makes the relevance pattern varies significantly. 
The representation-focused models are hard to capture such diverse matching patterns by relying on some simple interaction functions in the last layer. 
An alternative way is to employ \acp{PTM} to directly model complicated interaction patterns from low-level features.
Since existing \acp{PTM} pay more attention to the representation learning rather than the interaction learning in original pre-training objectives, researchers proposed different learning strategies to capture the query-document interactions by further pre-training \acp{PTM} in domain data.
According to the objective used in different pre-training models, we divide them into two categories: 1) Weak Supervised Learning; 2) Self-supervised Learning.


\subsection{Weak Supervised Learning}

Weak supervised learning aim to learn machine learning models on noisy data.
To be more specific, labels are automatically generated by other models not human beings.
And the learning objective of weak supervision is often the same as the objective of the downstream task, that is, the learning objective of weak supervision in IR is the ranking objective.
Once the models are pre-trained on the generated noisy data, they can also be fine-tuned with supervised training data on the target IR tasks~\citep{Dehghani2017NeuralRM,Luo2017TrainingDR,Zamani2018OnTT,Zamani2018NeuralQP,Zhang2020SelectiveWS}.

After the rise of NeuIR, researchers explored to pre-train a simple neural interaction model on weakly supervised data for ad-hoc retrieval to verify its effectiveness.
\citet{Dehghani2017NeuralRM} first investigated the weak supervised learning for IR. 
They train neural interaction models on billions of noisy training data automatically generated by BM25.
The input is query-document pairs and the model architecture is a simple feed-forward neural network.
Both pointwise learning and pairwise learning are studied under weak supervised setting.
Experiments showed the trained neural model using weak supervision can outperform BM25.
To study the reason, \citet{Zamani2018OnTT} theoretically analyzed weak supervision from the perspective of the risk minimization framework to verify its effectiveness.
Recently, \citet{Zhang2020SelectiveWS}  proposed a reinforcement weak supervision method with BERT, called ReInfoSelect.
ReInfoSelect trains a selector model to select some constructed anchor-document pairs for training the BERT-based ranker via reinforcement learning.
It takes the ranking performance (i.e., NDCG) as the reward.
Experiments showed the neural ranker trained by ReInfoSelect can match the effectiveness of neural rankers trained on private commercial search logs.

\subsection{Self-supervised Learning}

Self-supervised learning is somehow a blend of supervised learning and unsupervised learning~\citep{Liu2020SelfsupervisedLG,qiu2020pre}.
The basic idea of self-supervised learning is to predict any part of the input from other parts in some form, whose learning objective is not the same as the objective in the downstream tasks.
So the labels of training data are often from the data itself rather than the same as in a specific task, like relevance judgments in IR.
The learning paradigm of self-supervised learning is entirely the same as supervised learning.
Recent \acp{PTM} on pre-training interaction models such as BERT and StructBERT, aim to learn the \textbf{coherence} relationship between two sentences by predicting the sentence order.
Specifically, they usually take two sentences as input and pre-train the interaction model with Next Sentence Prediction (NSP) task or Sentence Order Prediction (SOP) task.
However, the coherence relationship quite diverges from relevance, which is the most important requirement of IR.
So, researchers designing \acp{PTM} tailored for IR mainly from the following two aspects:  1) Pre-training Objectives; 2) Model Architectures.



\paragraph{Pre-training Objectives}
Relevance is a vague notion in IR, so is there any other object to be a good proxy of relevance?
Inspired by the query likelihood model (QL)~\citep{ponte1998language}, \citet{Ma2021PROPPW} proposed a novel pre-training task named Representative wOrds Prediction (ROP) for ad-hoc retrieval, and the pre-trained model is called PROP.
QL assumes that the query is a piece of representative text generated from the ``ideal'' document~\citep{liu2005statistical}.
Thus, modeling \textbf{representativeness} may benefit to capture the relevance between the query and the document.
To verify this hypothesis, ROP samples pairs of word sets according to the multinomial unigram language model~\citep{Zhai2008StatisticalLM}, and then pre-trains the Transformer to predict the pairwise preference.
Experiments show PROP outperforms other pre-trained models like BERT and ICT on a variety of ad-hoc retrieval tasks.
Moreover, under both the zero-shot and few-shot settings, PROP can achieve surprising performance, and even outperform BM25 on Gov2 without fine-tuning.
\citet{Ma2021BPROPBP} further proposed B-PROP by leveraging BERT to replace the classical unigram language model for the ROP task construction.
Inspired by the divergence-from-randomness idea~\citep{amati2002probabilistic}, they proposed a contrastive method to leverage BERT's [CLS]-token attention to sample representative words.
Experiments show B-PROP performs better than PROP on the downstream document ranking datasets.
\citet{ma2021pre} proposed HARP with anchor texts and hyperlinks to replace the sampling method, as sampling may introduce noise to the data.
Experimental results show that HARP can perform better than PROP on MS-MARCO Document Ranking and TREC DL.
As most existing work adopts the two-stage training paradigm, models' off-the-shelf parameters can be largely updated in the fine-tuning process.
What knowledge on earth do these models have learned still remains under-investigated.
To this end, ~\citet{chen2022axiomatically} aimed to incorporate IR axioms into model pre-training and proposed a novel model named ARES.
They generated training samples with specific IR axioms or heuristics to guide the training of ARES.
Experimental results have shown the effectiveness of ARES, especially in low-resource scenarios where supervision data is limited.

\paragraph{Model Architectures}
Those works in Section~\ref{sec:p4ir_long} on designing \acp{PTM} for handling long texts can also be applied in pre-training interaction models.
There is less effort on designing new interaction model architectures for IR as the self-attention mechanism of the Transformer architecture does provide a solution to do interaction between texts.
In the fine-tuning phase, \citet{MacAvaney2020EfficientDR} proposed to block the attention flow between the query and the document at lower layers in a cross-encoder architecture.
Thus, they can pre-compute the document representations and accelerate the inference for re-ranking.

\section{Summary}

Fine-tuning the Transformer-based \acp{PTM} has dominated almost every component in IR due to its convenience and effectiveness in recent years. However, the performance improvement on different IR tasks was still limited since original pre-training objectives are designed to learn the language coherence, e.g., predicting the masked token or the sentence order~\citep{devlin2018bert}. To better leverage the pre-training paradigm for IR, there are two main lines of researches which concentrate effort on designing novel \acp{PTM} tailored for IR. 
The first one looks for novel pre-training objectives that better resemble IR requirements, e.g., the Inverse Cloze Task \citep{lee2019latent}, the Wiki Link Prediction \cite{chang2020pre}, and the representativeness of words prediction \citep{Ma2021PROPPW, Ma2021BPROPBP}. Though different learning objectives are introduced and claimed to be beneficial to IR tasks. However, it still remains unclear how good these learning objectives satisfy the IR requirements for lacking of theoretical basis. Moreover, some of the pre-training objectives is strongly related to the weak learning since both of them rely on heuristic rules of IR, and the difference between this two learning strategies has been less studied.
The second one focuses on designing new model architectures which aim to satisfy the heterogeneity structures in and between queries and documents, e.g., Longformer~\citep{Beltagy2020LongformerTL} and SEED~\citep{xiong2021seed}.
There are still very few works in this direction, and most of them have only made minor changes to original BERT model. This is due to the fact that the BERT model has been well trained on a very large-scale corpus, and a completely redesigned architecture leads to high model training cost. Moreover, it also requires in-depth analysis on the basis of the transformer architecture, and rethink the design criteria of architectures from the view of IR.
Finally, the fundamental question to the design of both pre-training objectives and architectures lies at the concept of the relevance in IR. Based on this view, it highlights the need for more systematic research concerning the definition of the relevance instead of heuristic hands-on learning objectives or model architectures.

%% file: Sections/7-resource.tex
\chapter{Resources of Pre-training Methods in IR}
\label{section:resource}

In this section, we sort out some popular data repositories which have potential for the pre-training and fine-tuning process of PTMs in IR.

\section{Datasets for Pre-Training}

\begin{sidewaystable}[htp]
\begin{tabular}{ccccc}
\toprule
\textbf{Dataset}   & \textbf{Source}  & \textbf{\#Docs} & \textbf{Language} &\textbf{Latest crawl date} \\ \hline
Books\tablefootnote{\url{https://github.com/huggingface/datasets/tree/master/datasets/bookcorpus}}     & Book     & 74M   & ENG      & 2015          \\
C4\tablefootnote{\url{https://github.com/huggingface/datasets/tree/master/datasets/c4}} & web extracted text & 0.3B & ENG & 2019 \\
Wikipedia\tablefootnote{\url{https://dumps.wikimedia.org/}} & Wiki text     & 10M    & Multi-lang      & monthly update                                    \\
RealNews\tablefootnote{\url{https://github.com/rowanz/grover/tree/master/realnews}} & News & 120GB & ENG & 2019 \\
Amazon\tablefootnote{\url{https://snap.stanford.edu/data/web-Amazon.html}} & reviews & 11GB & ENG & 2003 \\ \hline
WT10G\tablefootnote{\url{http://ir.dcs.gla.ac.uk/test_collections/wt10g.html}}     & web pages     & 1.7M   & ENG      & 1997          \\
GOV2\tablefootnote{\url{http://ir.dcs.gla.ac.uk/test_collections/access_to_data.html}}      & pages in .GOV & 25M    & ENG      & 2004           \\
CWP200T   & Chinese web pages     & 7B     & CHN      & 2015           \\
SogouT\tablefootnote{\url{http://www.sogou.com/labs/resource/t.php}}    & Sogou web pages     & 1.17B  & CHN      & 2016   \\ 
ClueWeb09\tablefootnote{\url{https://lemurproject.org/clueweb09/}} & web pages     & 1.04B  & Multi-lang      & 2009      \\
ClueWeb12\tablefootnote{\url{https://lemurproject.org/clueweb12/}} & web pages     & 0.73B  & ENG      & 2012           \\
MS MARCO\tablefootnote{\url{https://microsoft.github.io/msmarco/}}  & Bing web pages       & 3.2M   & ENG      & 2018            \\

\bottomrule
\end{tabular}
\caption{Public available datasets which are potential for pre-training tasks.}
\label{table:pretraining_dataset}
\end{sidewaystable}

As discussed in Section~\ref{section:PTM_for_information_retrieval}, pre-training objectives designed for IR are mostly based on a (or more) large-scale collection(s). We thus consider the collections for pre-training tasks in IR with the following properties:
\begin{itemize}[] 
	\item \textbf{Large collection size}: In a broad sense, collection size is a necessity for pre-training tasks in any deep learning fields. 
	\item \textbf{Structured documents}: The structures of a document include title, passages, sub-title, html structure, entity extractions, etc. These structures can be exploited in IR pre-training tasks to capture inter-page semantic relation. Moreover, hyperlinks between the pages(e.g., anchor-page linking and page-page linking) provide intra-page semantic relations, which can also be used in IR pre-training.
\end{itemize}

Specifically, we believe that the second property are not always necessary for IR pre-training tasks. But if a collection owns these properties, the collection might be better for IR pre-training tasks. Given the suggested properties of a IR pre-training dataset, we sort out some public available datasets which are potentially useful for pre-training tasks, as shown in Table~\ref{table:pretraining_dataset}. According to the closeness to the IR, we categorize existing datasets into general text corpus and IR related corpus:

\begin{itemize}[] 
	\item \textbf{General text corpus}: The general text corpus is widely used in NLP researches for different tasks in different domains. These datasets generally contain a large amount of documents and provide implications for the classic pre-training tasks, e.g., masked language modeling (MLM) and next sentence prediction (NSP). 
	\begin{itemize}
	    \item \textit{Books}: This dataset aims to align books with the corresponding movie releases by associating the visual information with descriptive text. The text conveys both visual content (how a character, an object or a scene looks like) as well as high-level semantics (what someone is thinking, feeling and how these states evolve through a story). 
        \item \textit{C4}: Colossal Clean Crawled Corpus (C4) is a dataset consisting of more than 300 GBs clean English text scraped from the web, which can be used to pretrain language models and word representations.
        \item \textit{Wikipedia}: Wikipedia is a large-scale collection containing all Wikimedia wikis in the form of wikitext source and metadata in XML structure.
It takes advantages in well-organized document structures, entity links, and rich information, which are suitable for pre-training tasks in IR.
        \item \textit{RealNews}: RealNews is a large-scale corpus containing news articles from Common Crawl. The documents are scraped from Common Crawl, limited to more than 5000 news domains indexed by Google News. News from Common Crawl dumps from December 2016 to March 2019 were used as training data; articles published in April 2019 were used for evaluation.
        \item \textit{Amazon Reviews}: This dataset consists of Amazon shopping reviews from amazon. The data spans a period of 18 years, including more than 35 million reviews up to March 2013. Reviews include user and product information, ratings, and a plaintext review.
        
    \end{itemize}
    
    \item \textbf{IR related corpus}: These kind of corpus contain documents which are similar to downstream IR tasks. Pre-training on these corpus can further minimize the gap between pre-training and downstream IR tasks, providing a better opportunity to achieve better ranking performance.
    \begin{itemize}
        \item \textit{WT10G}: WT10G (Web Track 10Gigabytes) was collected by CSIRO in Australia~\citep{chiang2005wt10g}. It is a crawl of web pages in 1997 and applied in many web-based experiments. The WT10G collection retains the properties of the 1997 web content which includes: the graph structure of web links, server size distribution, inclusion of inter-domain links and web pages on various subjects. The page content and hyperlinks in this dataset can be used in pre-training tasks by the methods discussed in Section~\ref{section:PTM_for_information_retrieval}.
        \item \textit{GOV2}: GOV2 is a crawl of .gov sites in the early of 2004 which includes html, text and the extracted text of pdf, word and postscript. The collection is about 426GB and contains 25 million documents. The large proportion of web pages has potential for pre-training tasks with text-based self-supervised learning objectives. 
        \item \textit{CWP200T, SogouT}: CWP200T and SogouT~\citep{luo2017sogout} are the web page collections in Chinese, which are provided by China Computer Federation (CCF) and Sogou search engine, respectively. Both collections are suitable for pre-training tasks in Chinese IR.
        \item \textit{Clubweb}: Clueweb is a large-scale web document collection provided by CMU. The full collection of Clueweb09 contains about 1 billion web pages in 10 languages which were collected in January and February 2009. 
Clueweb12 was further created based Clueweb09 with several data cleaning strategies.
Both datasets are widely used in IR and several tracks of the TREC conference.
        \item \textit{MS MARCO}: MS MARCO~\citep{Craswell2021MSMB} is a popular large-scale document collection consisting of 3.2 million available documents, which are from the Bing search engine. Besides, 1 million non-question queries are also included in this dataset for different retrieval tasks.
    \end{itemize}
    
\end{itemize}

For general text corpus, we believe there are a number of corpus which is not listed in our paper. We recommend readers to this link\footnote{\url{https://github.com/huggingface/datasets/tree/master/datasets}} to further explore the available datasets for pre-training tasks. And, the corpus with web pages mostly contain two important relations (i.e., inter-document (e.g., html structure) and intra-document (e.g., hyperlinks, anchor-page links) relations). These relations provides implications to design different pre-training objectives for IR tasks.

\section{Datasets for Fine-Tuning}
\label{section:data_finetune}
We sort out some datasets for downstream fine-tuning tasks. These tasks are categorized into document-oriented tasks and query-oriented tasks. The abbreviations of these tasks are further used in Table~\ref{table:data_finetune} as the potential tasks of different datasets. We introduce each specific task as follows:

\begin{itemize}
    \item \textbf{Document-oriented}
    \begin{itemize}
		\item \textit{First stage retrieval} (\textbf{FSR}): Retrieval stage from the full collection.
	    \item \textit{Ad-hoc ranking} (\textbf{AR}): Ranking a candidate list given a query.
	    \item \textit{Session search} (\textbf{SS}): Ranking a candidate list given a query and historical interactions.
	    \item \textit{Multi-modal ranking} (\textbf{MMR}): Given a query, rank the candidate list where each item contains multiple heterogeneous information such as text, picture and html structure.
	    \item \textit{Personalized Search} (\textbf{PS}): User-specific Ranking. 
	\end{itemize}
	\item \textbf{Query-oriented}
	\begin{itemize}
	    \item \textit{Query reformulation} (\textbf{QR}): Iiteratively modifying a query to improve the quality of search engine results in order to enhance user's search satisfaction.
		\item \textit{Query suggestion} (\textbf{QS}): Providing a suggestion which may be a reformulated query to better represent a user's search intent.
		\item \textit{Query clarification} (\textbf{QC}): Identifying user's search intent during a session.
	\end{itemize}
	\item \textbf{Others}
	\begin{itemize}
	    \item \textit{Document summarization} (\textbf{DS}): The process of shortening a document to create a subset (or a summary) that represents the most important information in this document.
		\item \textit{Snippet generation} (\textbf{SG}): Query-specific document summarization.
		\item \textit{Keyphrase extraction} (\textbf{KE}): It is also known as Keyword Extraction, which aims to automatically extract the most used and most important terms in a document.
	\end{itemize}
\end{itemize}

\begin{sidewaystable}[htp]
\footnotesize
\centering
\begin{tabular}{ccccc}
\toprule
\textbf{Dataset}          & \textbf{Subdata} & \textbf{Size}          & \textbf{Source} & \textbf{Potential Tasks}              \\ \hline
\multirow{2}{*}{Robust}   & Robust04 & 0.5M docs, 250 queries & \multirow{2}{*}{TREC Robust Track } & \multirow{2}{*}{FSR, AR, QR} \\
                          & Robust05 & 1M docs, 50 queries    &                   &                              \\ \hline
\multirow{2}{*}{TREC MQ}   & MQ2007 & 6.5K docs, 1.7K queries & \multirow{2}{*}{TREC Million Query track} & \multirow{2}{*}{FSR, AR, QR} \\
                          & MQ2008 & 1.4K docs, 784 queries    &                   &                              \\ \hline
\multirow{2}{*}{Clueweb}  & 09-CatB        & 50M docs, 150 queries                      & \multirow{2}{*}{Web pages} & \multirow{2}{*}{FSR, AR, QR, KE}            \\
                          & 12-CatB        & 50M docs                      &                   &                              \\ \hline
TREC web track & 99-2014 & See~\tablefootnote{\url{https://trec.nist.gov/data/webmain.html}} & TREC web track & FSR, AR, QR \\ 
TREC DL track & 2019-2021 & See~\tablefootnote{\url{https://microsoft.github.io/msmarco/TREC-Deep-Learning.html}} & TREC Deep Learning track & FSR, AR \\ 
AOL                       &   \textbackslash{}       & 6M queries                       &      AOL Query logs             &  AR,SS,PS,QR,QS                            \\ 
Sogou-QCL                 & \textbackslash{}         &    9M docs, 0.5M queries                    &        Sogou Query logs           &   AR, QR                            \\ 
Sogou-SRR                 &  \textbackslash{}        &  63K results, 6K queries                      &      Sogou Query logs             &  AR, MMR, QR                            \\ 
Tiangong-ST               &  \textbackslash{}        &   0.3M docs, 40K queries                     &      Sogou Query logs             &   AR, SS, QR, QS                           \\  
Qulac & \textbackslash{} & 10K question-answer pairs & TREC Web Track & AR, QR, QC \\ \hline

BEIR & 7 IR tasks & Vary from tasks & Wiki, Quaro, Twittter, News and etc. & FSR,AR, etc.\\ \hline
\multirow{2}{*}{MS MARCO} &   \multirow{2}{*}{2019-20}    &  \multirow{2}{*}{\begin{tabular}[c]{@{}c@{}}1M queries,\\  8.8M passages, 3.2M docs\end{tabular}}                       & \multirow{2}{*}{TREC Deep Learning Track} &  \multirow{2}{*}{FSR, AR, QR}           \\ 
 &         &                       & &            \\ \hline
TREC CAR        &   \textbackslash{}       &   30M paras, 2M queries                     &     TREC Complex answer retrieval & AR, QR, KE          \\
CNN / Daily Mail & \textbackslash{} & 0.3M docs & Human generated abstracts & DS\\
New York Times (NYT) &\textbackslash{} & 1.8M docs &  News
articles & DS \\ 
Debatepedia & \textbackslash{} & 1,303 debates &  Debate key points & SG, DS \\
DUC &2001-07 & 300 clusters, See~\tablefootnote{\url{https://duc.nist.gov/data.html}} &  Doc understanding conference & SG, DS \\
WIKIREF & \textbackslash{} & 0.3M samples & QFS benchmark & SG, DS\\
\bottomrule                  
\end{tabular}
\caption{Datasets for different downstream tasks in IR. Abbreviations in potential tasks are detailed in Section~\ref{section:data_finetune}.}
\label{table:data_finetune}
\end{sidewaystable}

The detailed description of each collection is as follows:
\begin{enumerate}
    \item Robust track~\citep{Voorhees2004Robust} is a classic ad-hoc retrieval task in TREC which focuses on poorly performing topics. The released annotated collection only includes 250 queries and 50 queries in Robust04 and Robust05, respectively. This collection is used for evaluation in most experimental settings.
    \item TREC Million Query (MQ) Track conducts an ad-hoc retrieval task over a large-scale collection of queries and documents. The final released dataset contains a four-level relevance judgement for each query-document pair.
    \item Clueweb is another large-scale web search dataset provided by CMU. The ``Category B'' data set consists of the English pages, which is roughly the first 50 million pages of the entire data set.
    \item TREC web track exploits the documents from Clueweb. The goal is to explore and evaluate specific aspects of Web retrieval, including traditional ad-hoc retrieval task, risk-sensitive task and diversity search task.
    \item TREC Deep Learning Track studies IR in a large training data regime. It contains two tasks: Passage ranking and document ranking; Two subtasks are included in each case: full ranking and reranking. Researchers usually take this dataset as an evaluation set by training a retrieval model on a large-scale dataset such as MSMARCO.
    \item AOL is a public available query log released by the internet company AOL. The collection contains the query session, anonymized user ids and clicked documents, which are suitable for ad-hoc ranking, session search ranking, personalized search ranking, query reformulation and suggestion.
    \item Sogou-QCL, Sogou-SRR (Search Result Relevance) and Tiangong-ST dataset were created from Sougou search engine to support research on IR. The Sogou-QCL collection consists of 537,366 queries, more than 9 million Chinese web pages, and five kinds of relevance labels assessed by click models. Meanwhile, the dataset also includes 2,000 queries with four-level human assessed relevance labels.
    \item The Sogou-SRR dataset consists of 6,338 queries and corresponding top 10 search results. Each search result contains the screenshot, title, snippet, HTML source code, parse tree, url as well as a four-grade relevance score (1-4) and the result type. The heterogeneous information provides opportunity for multi-modal ranking.
    \item Tiangong-ST provides 147,155 refined Web search sessions, 40,596 unique queries, 297,597 web pages, and six kinds of weak relevance labels assessed by click models. Different from Sogou-QCL and Sogou-SRR, the session information provided in this dataset is able to be used in session search ranking.
    \item Qulac was collected through crowdsourcing in terms of the topics in the TREC Web Track 2009-2012. It is a dataset on asking Questions for Lack of Clarity in open-domain information-seeking conversations. It contains 198 topics where each topic has recognized as either ``ambiguous'' or ``faceted''. The clarifying questions are collected based on each topic through crowdsourcing. Based on each topic-facet pair, the answers to each clarifying question are collected. The average number of facets per topic is 3.85 $\pm$ 1.05. The facets and topics in this collection can be used for query clarification task.
    \item BEIR (Benchmarking IR)~\citep{thakur2021beir} is a new heterogeneous benchmark containing different IR tasks. The benchmark contains 18 datasets covering 9 IR tasks (Fact Checking, Citation Prediction, Duplicate Question Retrieval, Argument Retrieval, News Retrieval, Question Answering, Tweet Retrieval, Biomedical IR, Entity Retrieval) from 17 different datasets.
    Through BEIR, it is possible to systematically study the zero-shot generalization capabilities of several neural retrieval methods.
    \item MS MARCO~\citep{Craswell2021MSMB} is a popular large-scale document collection which contains about 3.2 million available documents, which are from the Bing search engine. Besides, 1 million non-question queries are also included in this dataset for different retrieval tasks.
    \item The TREC Complex Answer Retrieval (CAR) track uses topics, outlines, and paragraphs that are extracted from English Wikipedia. Wikipedia articles are split into the outline of sections and the contained paragraphs. The complex topics are selected from articles on open information needs, i.e., not people, not organizations, not events, etc. It contains a passage task and an entity task, where the latter can be used in keyphrase extraction tasks.
    \item The CNN/Daily Mail dataset~\citep{see2017get} is a large-scale collection of news articles and further modified for summarization. It consists of more than 280,000 training samples and 11,490 test set samples. The documents in the training set have 29.74 sentences with 766 words on average while the summaries consist of 53 words and 3.72 sentences on average.
    \item New York Times (NYT)\footnote{\url{https://catalog.ldc.upenn.edu/LDC2008T19}} is a large-scale document summarization dataset. It contains well curated articles from~\textit{The New York Times} between 1987 and 2007. The summaries were written by library scientists, making it particularly useful as an extractive summarization dataset. 
    \item Debatepedia is collected from \textit{debatepedia.org}. It is an encyclopedia of pro and con arguments and quotes on critical debate topics. There are totally 663 debates in the corpus, which belong to 53 overlapping categories such as Politics, Law, Crime, Environment, Health, Morality, Religion, etc. The average number of queries per debate and documents per query is 5 and 4, respectively.
    \item The DUC dataset is a dataset for document summarization. In most experiments, it is used for testing only. It consists of 500 news articles, each of the article is paired with four human written summaries. In DUC2004, it consists of 50 clusters of Text REtrieval Conference (TREC) documents from the following collections: AP newswire, 1998-2000; New York Times newswire, 1998-2000; Xinhua News Agency (English version), 1996-2000. Each cluster contains on average 10 documents. For the details of other versions, please refer to here\footnote{\url{https://duc.nist.gov/data.html}}.
    \item WIKIREF is a large query-focused summarization dataset from Wikipedia which aims to generate summarization with a given query. It contains more than 280,000 examples.
\end{enumerate}

\section{Leaderboards}
In this section, we list several public leaderboards for researchers to understand the state-of-the-art methods in different tasks.

\begin{enumerate}
    \item MS MARCO (Passage retrieval and document retrieval task): \url{https://microsoft.github.io/msmarco/}
    \item DuReader (Machine Reading Comprehension task):\url{https://ai.baidu.com/broad/leaderboard?dataset=dureader}
    \item Robust04 (Document retrieval task): \url{https://paperswithcode.com/sota/ad-hoc-information-retrieval-on-trec-robust04 }
    \item CNN/Mail (Documents summarization task): \url{https://paperswithcode.com/sota/document-summarization-on-cnn-daily-mail}
    \item Baidu DuIE (Entity extraction task): \url{https://ai.baidu.com/broad/leaderboard?dataset=dureader}
    \item Benchmarking IR (BEIR) (Passage retrieval and document retrieval task): \url{https://github.com/UKPLab/beir}
\end{enumerate}

%% file: Sections/8-challenge.tex

\chapter{Challenges and Future Work}
\label{section:challenge}

In this chapter, we discuss current challenges and suggest some promising directions for pre-training methods researching in the IR field.

\section{New Objectives \& Architectures Tailored for IR}  \label{sec:Pre-training_for_IR}

Although the general-purpose pre-trained language models are suitable for learning the universal language knowledge, designing the pre-training and tuning methods that more closely resemble downstream tasks is admittedly a more efficient way to obtain better performance on specific tasks~\citep{zhang2020pegasus, Ke2020SentiLARESL}.
From the aspect of pre-training objectives, pre-training model architectures, and model tuning methods for IR, there have been some preliminary works, but we believe it deserves further exploration towards these directions.

\textbf{New Pre-Training Objectives.} 
As described in Section~\ref{section:PTM_for_information_retrieval}, there have been some pioneer studies~\citep{lee2019latent, chang2020pre, guu2020realm, Ma2021PROPPW, Ma2021BPROPBP, Liu2021PretrainedLM, ma2021pre} on the pre-training objectives tailored for IR. 
For example, \citet{lee2019latent} proposed to pre-train with a large-scale document collection with the Inverse Cloze Task (ICT) for retrieval tasks.
Besides ICT, \citet{chang2020pre} also proposed to capture the inner-page and inter-page semantic relations with Body First Selection (BFS) and Wiki Link Prediction (WLP) for passage retrieval in QA tasks.
For the re-ranking component, \citet{Ma2021PROPPW, Ma2021BPROPBP} proposed the Representative Words Prediction (ROP) objective for pre-training, which achieves significant improvement.
In addition to constructing pseudo query-document pairs from the raw text, some researches turned to relying on certain corpus structures.
For example, \citet{ma2021pre} proposed to leverage the large-scale hyperlinks and anchor texts for pre-training. Experimental results show that pre-training with four objectives based on the hyperlinks (i.e., RQP, QDM, RDP, and ACM) and the MLM objective jointly achieves state-of-the-art performance on two ad-hoc retrieval datasets.
On the whole, the underlying idea of all these pre-training objectives tailored for IR is to simulate the relevance relationship between queries and documents.
However, it is still in the preliminary stage to design more suitable pre-training objectives for IR.

\textbf{New Architectures.} 
Beyond designing new pre-training tasks for IR, another research line is to design novel architectures according to specific downstream tasks.
For example, towards the dual-encoder architecture for dense retrieval, \citet{gao2021your} argued that language models like BERT have a non-optimal attention structure to aggregate sophisticated information into a single dense representation for retrieval tasks. 
Based on these observations, they introduced a novel Transformer pre-training architecture, Condenser, to address structural readiness during pre-training. 
Experimental results show that Condenser yields stable improvement over standard LM and shows comparable performance to strong task-specific PTMs.
Similarly, in order to obtain better document embeddings for dense retrieval, ~\citet{xiong2021seed} presented a new auto-encoder architecture with restricted attention flexibility.
Based on this, the new architecture could create an information bottleneck in the auto-encoder and force the encoder to provide better document representations.
However, compared with attempts to investigate new pre-training objectives for IR, designing an ingenious pre-training model architecture which is suitable for IR tasks has not been well explored.

\textbf{Beyond Fine-Tuning.} 
Up to now, fine-tuning is the most dominant method to apply PTMs to downstream tasks, but it has some undesired limitations: (1) it performs poorly on some downstream tasks without enough supervision data to support fine-tuning; (2) it is inefficient to fine-tune parameters on every downstream task.
Recently, the emergence of GPT-3~\citep{GPT3} makes the prompt tuning~\citep{liu2021pre} attract more research attention. 
Prompt tuning needs to design discrete~\citep{petroni2019language, gao2020making} or continuous~\citep{liu2021gpt, Lester2021ThePO} prompts for specific downstream tasks.
For now, it is a promising way to reduce the computational cost of using pre-trained models for downstream tasks.
In fact, prompt tuning has achieved exciting results in some fields, such as information extraction~\citep{chen2021adaprompt, han2021ptr}, text classification~\citep{puri2019zero, Schick2021ExploitingCF}, and fact probing~\citep{petroni2019language, jiang2020can}. 
However, there has been no mature work on prompt tuning for IR tasks.
From another perspective, the design of most of existing PTMs is driven by the fine-tuning paradigm, but it is unclear whether the exploring of different PTMs will produce pre-trained models which are more effective when they are used with prompt tuning to solve IR tasks.

\section{Utilizing Multi-Source Data for Pre-training in IR}  \label{sec:Utilizing_Multi-Source_Data}
Developing PTMs based on multi-source heterogeneous data, including multi-lingual, multi-modal, and external knowledge, for IR is another promising direction.
On one hand, abundant data resources are vital significance for model pre-training, and on the other hand, incorporating extra data has great potential to enhance document representations for IR tasks.

\textbf{Multi-modal Pre-Training for IR.}  
Large-scale pre-training methods have been widely developed with diverse real-world modalities (e.g., text, image, audio, and video) and different practical applications.
In recent years, there has been an upsurging interest in cross-modal tasks, e.g., image-text retrieval~\citep{lee2018stacked, huo2021wenlan}, visual question answering~\citep{alberti2019fusion, antol2015vqa}, and image caption~\citep{vinyals2015show, johnson2016densecap}.
Meanwhile, PTMs based on cross modalities also have improved research interests, such as image-text~\citep{lu2019vilbert, li2019visualbert}, video-text~\citep{sun2019videobert}, or audio-text~\citep{chuang2019speechbert}.
Among the Vision-and-Language pre-training (VLP) research, most current works focus on the interaction of images and texts~\citep{li2019visualbert, su2019vl, lu2019vilbert, li2020oscar}, expecting to have a joint understanding of both to improve the performance on single-modal and multi-modal tasks.
Since 2019, many VLP models have been proposed and achieved great success for various downstream tasks.
Specially, \citet{cao2020behind} probed the pre-trained Vision-Language models over nine tasks in SentEval~\citep{conneau2018senteval}. Results show that the pre-trained model indeed encodes richer linguistic knowledge to enhance NLP tasks.
Similarly, the unified-modal pre-training architecture UNIMO~\citep{li2020unimo} models textual knowledge and visual knowledge in a unified semantic space and results in improved performance for NLP tasks.
However, most of these works are not evaluated on IR tasks.
Besides, although multi-modal PTMs has made great progress in recent years, \citet{cao2020behind} proved that the textual modality is more dominant than image during the multi-modal pre-training process. Based on this, the benefits of cross-modal learning are mainly reflected on image-based tasks. Thus, it is worth further exploring to design better vision-language pre-training objectives pointing at IR tasks.
On the other hand, utilizing more modalities (e.g., audio or video) and more data is another problem that needs to be further explored in the future.

\textbf{Multi-lingual Pre-Training for IR.}  
Despite the rapid progress in PTMs, most prior work has been exclusively on English, where large-scale annotations are easily available.
However, due to the cost and required dataset, pre-training large language models for each language is not practical.
Specially, the large-scale annotations are hard to obtain for low-resource languages.
Additionally, some empirical results show that training one model with several languages could get better performance on some tasks than training several monolingual models independently~\citep{lample2019cross, ni2021m3p}. 
Hence, training a language model based on multi-lingual data may be a good attempt for IR tasks.
In fact, some existing multi-lingual pre-trained models, such as mBERT~\citep{devlin2018bert}, XLM~\citep{lample2019cross}, and Unicoder~\citep{huang2019unicoder}, have shown their language transfer abilities over a wide range of tasks~\citep{wu2019beto}.
For example, \citet{shi2020cross} constructed the re-ranking model for non-English corpus  based on the mBERT, aiming to leverage the relevance information learned in English. They found that this significantly improves search quality for non-English retrieval.
However, most such works on multi-lingual PTMs focus on NLP tasks, and these multi-lingual PTMs are not well designed for cross-lingual tasks in IR.

\textbf{Knowledge-Enhanced Pre-Training for IR.}  
It is generally accepted that external knowledge, such as knowledge graphs and domain-specific data, can provide a good prior for model training.
Thus, introducing external knowledge into PTMs to get knowledge-enhanced representations for IR is another research line.
Based on knowledge graphs, there have been many explorations to integrate entity and relation embeddings or their alignments into pre-trained models training~\citep{zhang2019ernie, sun2019ernie, wang2021kepler}.
Different from structured knowledge, unstructured knowledge, e.g., the domain-specific data, is more abundant but also noisier. 
Several works~\citep{beltagy2019scibert, lee2020biobert} have attempted to further training the general pre-trained models on these data to get better performance for specific domains or tasks.
However, most of these efforts are not tailored for IR.
In the future, how to effectively model these knowledge for IR needs to be further explored.
On the other hand, all existing works store knowledge with model parameters implicitly.
How to model knowledge in a more interpretable way for downstream tasks has not been explored.

\section{End-to-End IR based on PTMs}  \label{sec:End-to-End_Learning}
Existing IR systems always follow a ``index-retrieve-rank" manner and separate three steps during training. 
However, this paradigm has some disadvantages in practical scenarios, which will produce sub-optimal performance.
Recently, the application of PTMs in the retrieval component makes the joint learning of multi-stages or end-to-end learning possible.

Technically, the index building process in retrieval systems based on the inverted index is hard to be trained jointly with the retrieval model. 
However, advances in PTMs-based retrieval models resulting in a shift from the inverted index towards the dense vector-based index makes the joint training possible.
In fact, there have been studies~\citep{zhang2021joint, zhan2021jointly, zhan2021learning} to explore the joint training of retrieval models and the index module. 
In this way, the index building can benefit from the relevance information between queries and documents directly.
In addition to the profits from the joint learning of index and retrieval, there have been works finding that it is beneficial to train retrievers and re-rankers in a correlated manner.
For example, the retriever can be improved by distilling knowledge from the re-ranker~\citep{ding2020rocketqa, hofstatter2020improving}, and the re-ranker can be improved with hard negatives generated from the retriever~\citep{gao2021rethink, huang2020embedding}.
Based on these observations, ~\citet{ren2021rocketqav2} proposed the dynamic listwise distillation to optimize two components jointly and contribute to the final ranking performance.
Nevertheless, these works are only preliminary attempts in this direction.
In fact, the joint learning of two components, i.e., retrieval and re-ranking, cannot be  implemented trivially and many problems have not been solved well.
Besides, researchers in this field have not ventured into the end-to-end learning of the whole pipeline, including indexing, retrieval, and re-ranking.

\section{Next Generation IR System: from Index-centric to Model-centric}  \label{sec:Next_Generation_IR_System}

Beyond the traditional multi-stage IR systems, the state-of-the-art pre-trained models with huge model size are capable of encoding more knowledge about the world, and based on this, they are probably able to generate results to information needs directly.
Thus, given the significant progress in PTMs, it is possible to set about the next generation of IR systems.

~\citet{metzler2021rethinking} proposed a vision to build model-based IR system based on the powerful pre-trained models.
Within the framework, the index is embedded into the model itself during the model training process, and retrieval and re-ranking components are implemented integrately with model inference. However, this work only gives a beautiful vision and vague framework.
Recently, ~\citet{Tay2022DSI} implemented this new IR paradigm based on the T5 model. The significant performance is achieved by training the model with indexing (i.e., documents to docids) and retrieval (i.e., queries to docids) in a multi-task setup.
At about the same time, ~\citet{Zhou2022DynamicRetriever} presented DynamicRetriever, which builds the model-based IR system based on BERT. They firstly fine-tuned the BERT-based dense retriever with query-document pairs, and then initialized the model parameters, especially the projection matrix with generated document embeddings. Finally, the model is further fine-tuned with query-docid pairs.
Nevertheless, these works are only preliminary explorations and there are still many deficiencies to be improved.
For example, how to build the semantics-based document identifications, and how to update the model when the document collection changes?
Besides, there are a number of challenges needing to be solved before the model-based IR system can be applied in practice.
At present, the capacity of existing pre-trained models is limited. For example, they do not have a real understanding of world knowledge, and it is challenging for them to develop the reasoning ability (e.g., arithmetic, logic, etc). Moreover, it is desiderative that the model-based IR system could be interpretable, debuggable and controllable. In fact, this is a core issue that all neural-based models need to address before they are applied.